\documentclass[aps,pra,10pt,twocolumn,notitlepage,superscriptaddress,dvipsnames,longbibliography, tightenlines,floatfix]{revtex4-1}
\pdfoutput=1
\usepackage{xcolor}
\usepackage{bm}
\usepackage{bbold}
\usepackage{mathtools}
\usepackage{graphicx}
\usepackage{paralist}
\usepackage[caption=false, subrefformat=parens, labelformat=parens]{subfig}
\usepackage{enumitem}
\usepackage{algorithm}
\usepackage{algpseudocode}
\usepackage{siunitx}

\usepackage{setspace}
\makeatletter
\def\BState{\State\hskip-\ALG@thistlm}
\makeatother
\setlist{leftmargin=.8cm}

\usepackage{soul} %for Strikethrough text
\usepackage{xcolor}
\definecolor{dark-red}{rgb}{0.4,0.15,0.15}
\definecolor{dark-blue}{rgb}{0.15,0.15,0.4}
\definecolor{medium-blue}{rgb}{0,0,0.5}
\usepackage{hyperref}
\usepackage[all]{hypcap}%fix problems with figures and tables of the 'hyperref' package
\usepackage{cleveref}% Package hperref works wrongly within the 'subequations' environment without this.
\hypersetup{
    colorlinks, linkcolor={dark-red},
    citecolor={dark-blue}, urlcolor={medium-blue}
} %These setups replace the ugly and distracting red and green boxes of the 'hyperref' package with colored texts.
\newcommand{\ssp}{\hspace{0.4pt}}
\newcommand{\norm}[1]{\lvert #1 \rvert}

\newcommand{\ket}[1]{\lvert\, #1\, \rangle}

\newcommand{\bra}[1]{\langle\, #1\, \rvert}

\newcommand{\anticommute}[2]{{\{\, #1,\, #2 \,\}}}
\newcommand{\half}{\frac{1}{2}}

\newcommand{\hc}{\mathrm{h.c.}}

\newcommand{\ea}{{\it et al.}}

\newcommand{\code}[3]{{$[[\ssp#1,\ssp #2,\ssp #3\ssp]]$}}
\newcommand{\nn}{n}

\newcommand{\ff}{f}

\newcommand{\cc}{c}

\newcommand{\XX}{X}
\newcommand{\YY}{Y}
\newcommand{\ZZ}{Z}

\newcommand{\length}{\ell}

\newcommand{\parity}{P}
\newcommand{\Majo}{\mathcal{M}}

\newcommand{\zz}{z}

\newcommand{\BKSF}{BKSF}
\newcommand{\MLSC}{MLSC}
\newcommand{\GFT}{$\mathrm{GF}(2)$}

\graphicspath{{./Figures/}}

\begin{document}

\title{Majorana loop stabilizer codes for error correction in fermionic quantum simulations}

\date{\today}

\author{Zhang Jiang}
\email{qzj@google.com}
\affiliation{Google Research, Venice, CA 90291}

\author{Jarrod McClean}
\affiliation{Google Research, Venice, CA 90291}

\author{Ryan Babbush}
\affiliation{Google Research, Venice, CA 90291}

\author{Hartmut Neven}
\affiliation{Google Research, Venice, CA 90291}

\begin{abstract}
Fermion-to-qubit mappings that preserve geometric locality are especially useful for simulating lattice fermion models (e.g., the Hubbard model) on a quantum computer.  They avoid the overhead associated with geometric non-local parity terms in mappings such as the Jordan-Wigner transformation and the Bravyi-Kitaev transformation.  As a result, they often provide quantum circuits with lower depth and gate complexity.  In such encodings, fermionic states are encoded in the common $+1$ eigenspace of a set of stabilizers, akin to stabilizer quantum error-correcting codes.  Here, we discuss several known geometric locality-preserving mappings and their abilities to correct/detect single-qubit errors.  We introduce a geometric locality-preserving map, whose stabilizers correspond to products of Majorana operators on closed paths of the fermionic hopping graph.  We show that our code, which we refer to as the Majorana loop stabilizer code (MLSC) can correct all single-qubit errors on a 2-dimensional square lattice, while previous geometric locality-preserving codes can only detect single-qubit errors on the same lattice.  Compared to existing codes, the MLSC maps the relevant fermionic operators to lower-weight qubit operators despite having higher code distance.  Going beyond lattice models, we demonstrate that the MLSC is compatible with state-of-the-art algorithms for simulating quantum chemistry, and can offer those simulations the same error-correction properties without additional asymptotic overhead.  These properties make the MLSC a promising candidate for error-mitigated quantum simulations of fermions on near-term devices.   
\end{abstract}

\maketitle

\section{Introduction}
\label{sec:intro}

We are closer to realizing the potential of quantum computation~\cite{feynman_simulating_1982,lloyd_universal_1996,georgescu_quantum_2014} with the recent rapid advances in quantum computing devices such as trapped ions~\cite{cirac_quantum_1995,kielpinski_architecture_2002,haffner_quantum_2008} and superconducting qubits~\cite{devoret_superconducting_2004,wendin_quantum_2017}.  While performing fault-tolerant quantum computation~\cite{Fowler_Mariantoni_Martinis_Cleland_2012} is widely regarded as the ultimate goal, the daunting overheads prevent it from being implemented in the immediate future~\cite{devitt_requirements_2013,ogorman_quantum_2017}.  In the meantime, error-mitigation schemes are likely to be an essential component of near-term quantum algorithms~\cite{gottesman_quantum_2016,roffe_protecting_2018,McClean:2017a,Colless:2018,temme_error_2017,harper_fault-tolerant_2019,mcardle_error-mitigated_2019,bonet-monroig_low-cost_2018}.

Fermionic systems must be mapped to spin systems before they can be simulated on a digital quantum computer~\cite{ortiz_quantum_2001}.  When the spatial dimension is greater than one, mappings such as the Jordan-Wigner (JW) transformation introduce non-local parity terms that act on a large number of qubits.  These parity terms add considerable overhead to quantum simulations of local fermionic systems.  Clever circuit compilation methods have been introduced to reduce the overhead of the parity terms on digital quantum computers with all-to-all connections~\cite{hastings_improving_2015,wecker_solving_2015}.  On the other hand, mappings such as the Bravyi-Kitaev (BK) transformation~\cite{bravyi_fermionic_2002} reduces such overheads considerably.  However, the resulting qubit operators of the BK transformation are still geometrically non-local.  This problem becomes prominent on near-term quantum devices where only short-range qubit-qubit interactions are available~\cite{houck_on-chip_2012,barends_digitized_2016}.  One way to tackle this problem is to use the fermionic SWAP network to bring together fermionic modes encoded far apart in the JW transformation~\cite{bravyi_fermionic_2002}, which offers asymptotically optimal solution to simulating quantum chemistry problems in terms of circuit depth~\cite{babbush_low-depth_2018,kivlichan_quantum_2018,Motta2018}.  For other problems, such as the two-dimensional (2D) fermionic Fourier transformation, the fermionic SWAP network is not optimal; this transformation can be implemented with a circuit depth that is quadratically lower in the problem size~\cite{jiang_quantum_2018}.

Fermion-to-qubit mappings that preserve geometric locality avoid the overhead of the fermionic SWAP networks, which is suitable for simulating fermionic systems with short range interactions.  Bravyi and Kitaev~\cite{bravyi_fermionic_2002} described a way to map fermionic hopping Hamiltonians defined on a graph $G$ to qubit operators on the edges of $G$ (or more precisely, the vertex-to-edge dual graph of $G$); two qubits can interact with each other if and only if their corresponding edges in $G$ share a common vertex.  A stabilizer operator (also called gauge operator in the literature~\cite{Ball_2005, Chen_Kapustin_Radicevic_2018}) is introduced for each closed path in the graph $G$, and the fermionic state is encoded in the common $+1$ eigenspace of these stabilizers.  This mapping is called the Bravyi-Kitaev superfast (\BKSF) transformation~\cite{whitfield_local_2016} or ``superfast'' encoding~\cite{setia_superfast_2019}, referring to ``superfast simulation of fermions on a graph."  This is distinct from the more widely known Bravyi-Kitaev transformation introduced in the same paper~\cite{bravyi_fermionic_2002}; the BK transformation does not preserve geometry locality and requires no more qubits than the number of fermionic modes.  

% Fermionic states remain the same after an excitation travel through a closed path in the hopping graph $G$.  In contrast, the qubit state picks up a parity term (product of Pauli-$Z$ operators along the path) during the same process.  In geometric locality-preserving maps, one extra conserved quantity has to be introduced for each independent closed path in $G$.  The number of qubits required is $N_Q = N_V + N_L - 1$, where $N_V$ ($N_L$) is the number of vertices (independent loops) of $G$ and the minus one comes from conservation of the total parity.  According to Euler's equation $N_E = N_V + N_L - 1$, the number of qubits $N_Q$ equals to the number of edges $N_E$ of $G$.  Mapping fermions to qubits on the vertex-to-edge dual graph allows one to introduce extra conserved quantities corresponding to the fermionic parities for closed paths while keeping the degrees of freedom of the original system unchanged.

Geometric locality-preserving maps are closely related to lattice gauge theories.  Levin and Wen~\cite{levin_fermions_2003,wen_quantum_2003} showed that particular fermionic lattice models can be described in terms of gauge fields in lattice spin models.  Ball~\cite{ball_fermions_2005} extended the Levin-Wen result to general fermionic hopping Hamiltonians by introducing auxiliary Majorana modes; Verstraete and Cirac~\cite{verstraete_mapping_2005} discussed a similar map for fermions on $d$-dimensional square lattices, where $d-1$ auxiliary qubits are introduced for each fermionic mode.  This approach is called the Ball-Verstraete-Cirac (BVC) transformation or the auxiliary fermion approach.  Chen~\emph{et al.} introduced a map similar to the one by \BKSF\ with examples from free fermions on square and honeycomb lattices to the Hubbard model~\cite{Chen_Kapustin_Radicevic_2018}.  Recently, Steudtner and Wehner~\cite{steudtner_quantum_2019} also constructed a class of locality-preserving fermion-to-qubit mappings by concatenating the JW transformation with some quantum codes.  Applications of locality-preserving maps to quantum simulation were discussed in~\cite{whitfield_local_2016,setia_bravyi-kitaev_2018}.

We expect error-mitigation methods to be useful in near-term quantum algorithms, such as the variational quantum eigensolver (VQE)~\cite{mcclean_theory_2016}.  A crucial step in either VQE or phase estimation is to prepare the initial state, which is often taken to be the ground state of a mean-field approach such as Hartree-Fock.  In chemistry, the difference between the mean-field energy and the exact energy is referred to as the ``correlation energy'', and despite determining essential elements of chemical bonding, it often only amounts to a few percent of the total energy~\cite{Helgaker2002,hammond1994monte}.  Physically, this is because a large fraction of the energy comes from core electrons that contribute very little directly to chemical bonding.  However, because of the energy scale difference, even small errors in this part of the calculation can overwhelm any refinements a quantum computer can provide.  Therefore, even a 1\% error in the preparation of the mean-field state could render any correlations added by the rest of the circuit fruitless.  One can drastically reduce the error rate in state preparation by using an error-detecting code and post selecting the measurement outcomes~\cite{mcardle_error-mitigated_2019,bonet-monroig_low-cost_2018,McClean:2017a,Colless:2018}.  An error detecting code can only detect errors but not correct them; simply knowing that an error has occurred, is much less useful than also being able to fix it.  Furthermore, error-detecting codes can also be used for quantum error suppression by creating an energy gap between the code subspace and the orthogonal subspace~\cite{young_error_2013,bookatz_error_2015,jiang_non-commuting_2017}. 

Recently, Setia~\ea~\cite{setia_superfast_2019} discussed a generalized version of the \BKSF\ transformation and their relations to error-correcting codes.  A no-go theorem was also proven in~\cite{setia_superfast_2019} showing that the \BKSF\ defined on graphs with vertex degree $d\leq 6$ cannot correct all single-qubit errors.  Here, we introduce a type of geometric locality-preserving mappings based on the \BKSF, called Majorana loop stabilizer code (MLSC).  The MLSC on a square lattice (vertex degree 4) can correct all single-qubit errors while the \BKSF\ can only detect single-qubit errors.  It is usually harder to implement codes with higher distances due to the higher weights of the logical operators.  One the contrary, the logical operators relevant to fermionic simulations in a MLSC typically have lower weights than those in \BKSF, making it easier to implement.  The stabilizers in the MLSC, however, usually have higher weights than those in the BKSF.  This is less of a problem in near-term experiments, where stabilizer measurements should only be implemented sparsely due to the large overhead (time and fidelity) associated with measurements and classical feedback loops.  The complication of syndrome measurements can also be mitigated using a method recently introduced by the authors in~\cite{mcclean_decoding_2020}, where syndrome detection is achieved by classical post processing; the drawback of this approach is that it can only be applied to the very last round of stabilizer measurements.  Even without the error-correction properties, the circuits associated with simulating lattice models using MLSC are as small as the best circuits in any prior work.  The MLSC is reasonably practical to implement in experiment, and perhaps more importantly, develop schemes on encoded quantum simulations of fermions that offer an advantage over unencoded approaches. 

% Standard mappings like Jordan-Wigner or regular Bravyi-Kitaev map fermionic operators to qubit operators with weight $O(N)$ or $O(\log N)$, but the terms are geometrically non-local. Transformations that conserve geometric locality, such as ours, BKSF and Verstraete-Cirac (VC) introduce some overhead in needing more qubits (approximately twice as many qubits are required for the 2D Hubbard model); however, for near-term devices with limited connectivity the geometric locality is very attractive because it asymptotically reduces the number of operations that must be performed.    

Previously, mappings which preserve geometric locality had only seemed practical as a strategy for simulating lattice fermion models.  While one could apply BKSF and MLSC directly to the Hamiltonians of arbitrary fermion models, such as those encountered in quantum chemistry, doing that is highly impractical as the number of qubits required would scale like the number of terms in a fully connected one-body fermion operator, which scales generically $O(N^2)$ for arbitrary electronic structure models with $N$ orbitals). However, here we describe a simple technique which allows one to perform quantum chemistry simulations within the MLSC (or BKSF) so that one can benefit from the error-correction properties of the MLSC, without any asymptotic increase in the spatial complexity or gate complexity. The essential insight is that one can implement certain Hamiltonian simulation algorithms within these codes without explicitly encoding the actual Hamiltonian being simulated. Rather than encoding the quantum chemistry Hamiltonian, we propose that one use the MLSC to encode a two-dimensional lattice of fermions. Then, using techniques from \cite{Motta2018} one can decompose the quantum chemistry Hamiltonian into a representation that can be efficiently simulated with only nearest-neighbor gates~\cite{kivlichan_quantum_2018}. Because only nearest-neighbor fermionic interactions are required, the MLSC on a lattice is sufficient to encode all of the required gates. This makes our techniques as applicable for chemistry as for lattice models.

In Sec.~\ref{sec:majorana}, we review some properties of the Majorana fermions that are crucial to the rest of paper.  In Sec.~\ref{sec:bksf}, we review the \BKSF\ and how it can be used to detect single-qubit errors. In Sec.~\ref{sec:mlsc}, we introduce the Majorana loop stabilizer code and discuss how to use them to correct single-qubit errors.  In Sec.~\ref{sec:obc}, we discuss how to construct MLSC with open boundary conditions.  In Sec.~\ref{sec:encoding}, we discuss how to encode a single Slater determinant with locality preserving mappings.  In Sec.~\ref{sec:chemistry}, we discuss how these encodings may be useful for simulating fermion models with high degree connectivity, such as those encountered in quantum chemistry. In Sec.~\ref{sec:conclusion}, we conclude the paper.  In App.~\ref{sec:error-correcting}, we briefly review quantum error-correcting codes.  In Apps.~\ref{sec:auxiliary_fermion} and \ref{sec:combine_unit_cells}, we discuss how to construct quantum error-correcting/detecting codes from the auxiliary fermion approaches.  In Apps.~\ref{sec:general_mlsc}, we discuss how to construct general MLSCs. 

\section{Majorana operators}
\label{sec:majorana}

In this section, we review some properties of the Majorana operators which will be used to construct the mappings in the later sections.  The Majorana operators have the advantage of being Hermitian and unitary at the same time (self-inverse).  They also allow for treating all the fermionic operators on an equal footing.  The single-mode Majorana operators are defined as linear combinations of the fermionic ladder operators
\begin{align}\label{eq:majo_ops}
 \ff_{2k} = \cc_k^\dagger + \cc_k\,,\quad   \ff_{2k+1} = i\ssp ( \cc_k^\dagger - \cc_k)\,,
\end{align}
where $\cc_k^\dag$ and $\cc_k$ are the fermionic creation and annihilation operators, respectively.  They satisfy the simple relations  
\begin{align}\label{eq:anti_commut}
 \ff_k^\dagger = \ff_k\,,\quad \anticommute{\ff_j}{\ff_k} = 2\delta_{j,k}\,,
\end{align}
where $\{\cdot\,,\,\cdot\}$ is the anticommutator and $\delta$ the Kronecker delta function. 

The single-mode Majorana operators $\ff_0,\ff_1,\ldots, \ff_{2\nn-1}$, together with the phase factor $i$, generate a group $\Majo_{2\nn}$.  An arbitrary element in $\Majo_{2\nn}$ takes the form 
\begin{align}\label{eq:multi_majo}
 \varphi\, \ff_A\,,\qquad  \ff_A  = \prod_{k\in A} \ff_k\,,
\end{align}
where $\varphi \in\{\pm 1, \pm i\}$ is an overall phase factor.  The set $A \subseteq \{0,1,\ldots, 2\nn-1\}$ is the support of the Majorana operator.  The order of the operators in the product $\ff_A$ matters, and we follow the convention that the index $k$ increases monotonically from left to right.  The weight of a Majorana operator equals the number of fermionic modes that it acts non-trivially on, i.e., the weight of $\ff_A$ is $\norm{A}$.  Two Majorana operators $\ff_A$ and $\ff_B$ either commute or anticommute, 
\begin{align}\label{eq:commut_majo}
 \ff_A \ff_B = (-1)^{\norm{A}\cdot \norm{B} +\norm{A\cap B}}\, \ff_B \ff_A \,.
\end{align}
The commutation relationship is determined by the parity of the overlap $\norm{A\cap B}$ when either $\norm{A}$ or $\norm{B}$ is even. 
% Inversely, we have
% \begin{align}\label{eq:Majo_inverse}
%  \cc_j^\dagger =\half \big(\ff_{2j} - i\ff_{2j+1}\big)\,,\quad \cc_j =\half \big(\ff_{2j} + i\ff_{2j+1}\big)\,.
% \end{align} 

In terms of Majorana operators, the fermionic occupation operator reads
\begin{align}\label{eq:fermi_occu}
 \cc_k^\dagger\cc_{k} = \half \Big(\openone+i\ff_{2k}\ff_{2k+1}\Big)\,,
\end{align}
where $\openone$ is the identity operator.  The fermionic hopping term takes the form
\begin{align}\label{eq:fermi_hop}
 \cc_{j}^\dagger\cc_k + \hc = \frac{i}{2}\Big(\ff_{2j}\ff_{2k+1}-\ff_{2j+1}\ff_{2k}\Big)\,.
\end{align}
For now we will focus on problems where fermions can only hop between neighboring vertices of a undirected graph $G(V,E)$, with vertices $V$ and edges $E$ (later we will explain how these techniques can be used to simulate models with arbitrary interactions, such as those in quantum chemistry).  Consider the following set of quadratic Majorana operators 
\begin{subequations}\label{eq:quadratic_majo_generators}
\begin{align}
 &\eta_k = i\ff_{2k}\ff_{2k+1} \,,&&\text{for each vertex  $k \in V$,}\\[3pt]
 &\xi_{jk} = i\ff_{2j}\ff_{2k}\,, &&\text{for each edge $(j,\,k)\in E$.}
\end{align}
\end{subequations}
The occupation term~\eqref{eq:fermi_occu} and the hopping term~\eqref{eq:fermi_hop} can be expressed as
\begin{subequations}
\begin{gather}
\cc_k^\dagger\cc_{k} = \half \big(\openone+\eta_k\big)\,,\label{eq:occupation}\\
\cc^\dagger_j \cc_k + \hc = -\frac{i}{2}\big(\xi_{jk} \eta_k + \eta_j\xi_{jk}\big)\,.\label{eq:hopping}
\end{gather}
\end{subequations}
The operators in Eq.~\eqref{eq:quadratic_majo_generators} generate a group $\Majo_{2\nn}^\mathrm{even}$, consisting of Majorana operators of even weights.  Any physical fermionic Hamiltonian can be written as a sum of elements in the group $\Majo_{2\nn}^\mathrm{even}$.

The quadratic Majorana generators in Eq.~\eqref{eq:quadratic_majo_generators} are both Hermitian and unitary. They satisfy the commutation relations~\cite{bravyi_fermionic_2002}
\begin{subequations}\label{eq:commu_quadratic_majo}
\begin{gather}
\eta_j \eta_k = \eta_k\eta_j\,,\\
\eta_l\ssp\xi_{jk} = (-1)^{\delta_{jl}+\delta_{kl}}\xi_{jk}\eta_l\,,\\
\xi_{lm}\ssp\xi_{jk} = (-1)^{\delta_{jl}+\delta_{kl}+\delta_{jm}+\delta_{km}}\xi_{jk}\ssp\xi_{lm} \,.\label{eq:commu_quadratic_majo_c}
\end{gather}
\end{subequations}
In addition to Eq.~\eqref{eq:commu_quadratic_majo}, we introduce the following condition on any closed path in the graph $G$ to close the algebra,
\begin{align}\label{eq:closed_path_relation}
(-i)^\length\,\xi_{k_0 k_1} \xi_{k_1 k_2}\cdots \xi_{k_{\length-1} k_0}  = \openone\,,
\end{align}
where $\length$ is the length of the path. This relation implies that the generators in Eq.~\eqref{eq:commu_quadratic_majo} are not independent. The total number of independent closed paths is $N_E-N_V+1$, where $N_V$ ($N_E$) is the total number of vertices (edges) of the graph $G$.   In the next section, we will see that the condition~\eqref{eq:closed_path_relation} is not automatically satisfied when the fermion operators are mapped to qubit operators.

\section{The Bravyi-Kitaev superfast transformation}
\label{sec:bksf}

In this section, we review the fermion-to-qubit mapping introduced by Bravyi and Kitaev that conserves geometry locality~\cite{bravyi_fermionic_2002}.  This mapping was called Bravyi-Kitaev superfast (\BKSF) transformation~\cite{whitfield_local_2016} or superfast encoding~\cite{setia_superfast_2019}, referring to ``superfast simulation of fermions on a graph."  Concerning the Hubbard model, it was shown that the \BKSF\ has an advantage over the JW transformation and the BK transformation in terms of total number of gates~\cite{havlicek_operator_2017}.  Here, we also show how to use the \BKSF\ to detect (but not correct) all single-qubit errors on a 2D square lattice.

\begin{figure}[htb]\label{fig:bk_graph}
\centering
\includegraphics[width=0.28\textwidth]{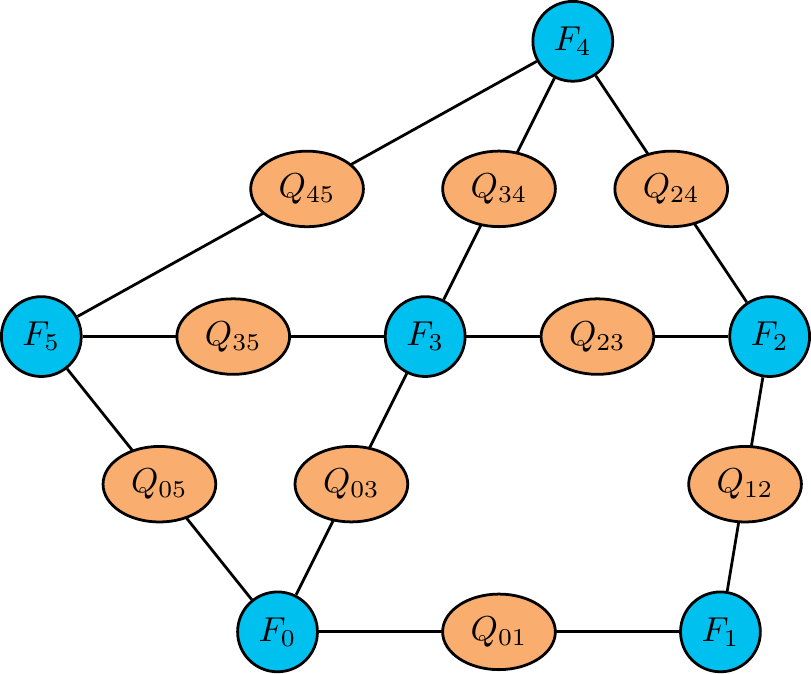}
\caption{\BKSF\ on a graph: the vertices (blue circles) of the graph represent fermionic modes; fermions can hop between neighboring vertices.  One qubit (orange ellipses) is introduced for each edge of the graph.  The fermionic state is mapped to a subspace of the qubits. }
\end{figure}

Bravyi and Kitaev considered fermionic hopping problem defined on a graphs $G$, where fermions can hop between neighboring vertices.  One qubit is introduced for each edge of the graph $G$, and the fermionic state is encoded in a subspace of the qubits, see Fig.~\ref{fig:bk_graph}.  
The goal is to find qubit operators that share the same algebra as the fermionic operators, e.g., Eqs.~\eqref{eq:commu_quadratic_majo} and \eqref{eq:closed_path_relation}.  To achieve that, we first choose an arbitrary ordering of the incident edges for each vertex of $G$, e.g., $(l,k) < (j,k)$ means that the edge $(l,k)$ is placed before $(j, k)$ among all the incident edges of the vertex $k$.  
%These orders can be arbitrarily chosen, but one has to be consistent with them.  
The quadratic Majorana operators in Eq.~\eqref{eq:commu_quadratic_majo} can be mapped to the qubit operators by
\begin{gather}
\eta_k \mapsto \tilde\eta_k = \prod_{j:\,(j,k)\in E} Z_{j  k}\,,\label{eq:eta_qubit}\\
\xi_{jk}\mapsto \tilde\xi_{jk}= \epsilon_{jk} X_{jk}\!\prod_{\substack{l:\,(l,k) < (j,k);\\ (l,k)\in E}}\!\! Z_{l k}\!\!\prod_{\substack{m:\, (m,j) < (k, j);\\ (m,j)\in E}} \!\!Z_{m j}\,,\label{eq:xi_qubit}
\end{gather}
where the antisymmetric coefficients $\epsilon_{jk}=-\epsilon_{kj}=\pm 1$ can be chosen arbitrarily for each edge $(j,\, k)$.  The qubit operators $\tilde\eta_k$ and $\tilde\xi_{jk}$ satisfy the same commutation relations as those of the fermionic operators in Eq.~\eqref{eq:commu_quadratic_majo}.

The qubit operator corresponding to the conserved quantity in Eq.~\eqref{eq:closed_path_relation} is
\begin{align}\label{eq:closed_path_relation_qubit}
S_{k_0\ssp k_1\ssp \cdots\ssp k_{\length-1}} = (-i)^\length\, \tilde\xi_{k_0 k_1}\ssp \tilde\xi_{k_1 k_2}\cdots \tilde\xi_{k_{\length-1} k_0}  \,,
\end{align}
which remains unchanged under any cyclic permutation of the indices $k_0, k_1,\ldots, k_{\length-1}$.  These loop operators form an Abelian group, and the fermionic state is encoded in their common $+1$ eigenspace.  We refer these operators as stabilizer operators.  They are also called gauge operators in the literature, not to be confused with the noncommuting gauge operators in subsystem quantum error-correcting codes~\cite{bacon_operator_2006}.  As a consequence of the commutation relations~\eqref{eq:commu_quadratic_majo}, the stabilizers commute with all the logical operators $\tilde\eta_k$ and $\tilde\xi_{jk}$.  They also commute with each other, and their common $+1$ eigenspace define the code subspace $\mathcal C$,
\begin{align}
\ket{\psi}\in \mathcal C\,,\; \text{if and only if}\;\,S_{k_0\ssp k_1\ssp \cdots\ssp k_{\length-1}}\ket{\psi}= \ket{\psi}\,,
\end{align}
for all closed paths $(k_0,k_1,\ldots, k_{\length-1}, k_0)$ in the graph $G$.  Hence, the restriction of the closed loop stabilizers to the code space act as the identity, as in the case of the Majorana operators.

The mapping~\eqref{eq:eta_qubit} artificially introduces an extra conserved quantity,
\begin{align}\label{eq:even_parity}
\prod_{k=0}^{N_V-1} \tilde\eta_k = \openone\,,
\end{align}
which corresponds to the even-parity subspace of the original fermionic system.   To simulate the odd-parity subspace, one can change the sign in the mapping~\eqref{eq:eta_qubit} for a particular value of $k$, say $k=0$,
\begin{align}
 \tilde\eta_k \mapsto (-1)^{\delta_{k,0}}\!\! \prod_{j:\,(j,k)\in E} Z_{j  k}\,.
\end{align}

The stabilizer operators in the \BKSF\ can be used for quantum error detection/correction if implemented properly.  This idea was also discussed recently in~\cite{setia_bravyi-kitaev_2018}, where a no-go theorem was proved: the \BKSF\ cannot correct all single-qubit errors if $d\leq 6$, where $d$ is the vertex degree of the graph $G$.  Here, we show that \BKSF\ can detect all single-qubit errors on a 2D square lattice by properly choosing the orders of the incident edges. 

\begin{figure}[htb]\label{fig:bk_lattice_c}
\centering
\includegraphics[width=0.31\textwidth]{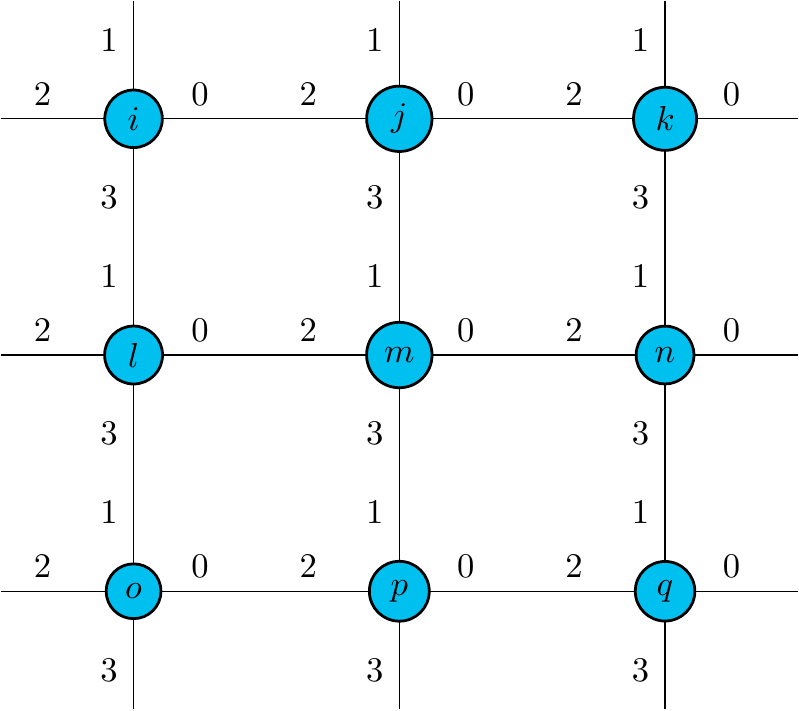}
\caption{\BKSF\ on a 2D lattice with a specific ordering of the incident edges.  This code can detect all single-qubit errors.}
\end{figure}

\begin{figure}[htb]
\label{fig:bksf_stab}
\centering
\includegraphics[width=2.7cm]{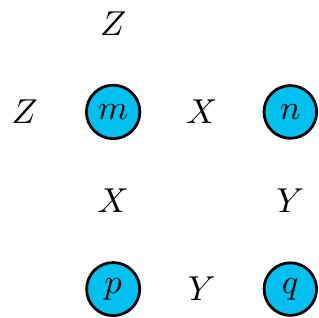}
\caption{Plaquette stabilizer for the BKSF code given the vertex and edge ordering in Fig \ref{fig:bk_lattice_c}, up to an overall minus sign.  From the chosen ordering, the above figure specifies the stabilizer that results from any generating plaquette in the graph.  In the case of the infinite or periodic lattice, it is uniform.  In the case of a finite lattice, the structure stays roughly the same; however, the dangling $Z$ operators are removed in the case that there is no edge in the graph, and the adjacent edges may change from $X$ to $Y$. \label{fig:bk_plaquette}}
\end{figure}

Consider the \BKSF\ encoding of spinless fermionic modes on a 2D lattice, see Fig.~\ref{fig:bk_lattice_c}; the ordering of the edges associated with a vertex is indicated by the numbers next to it.
We use the convention that $\epsilon_{jk} = 1$ if the site $k$ is to the right or below the site $j$; under this convention, we have $\epsilon_{k_0 k_1} \epsilon_{k_1 k_2}\cdots \epsilon_{k_{\length-1} k_0} = 1$ for any closed path.  The closed paths containing the smallest number of edges are plaquettes.  The stabilizer operator corresponding to the plaquette $(m,n,q,p,m)$ is the product of the corresponding edge operators, see Fig~\ref{fig:bksf_stab},
\begin{align}
 S_{mnqp} &= \tilde\xi_{m n} \,\tilde\xi_{nq}\, \tilde\xi_{qp}\, \tilde\xi_{pm}\nonumber\\
 &= X_{mn} X_{nq}Z_{nq} X_{qp}Z_{qp} X_{pm} Z_{jm} Z_{lm} \nonumber\\
 &= - X_{mn} Y_{nq} Y_{qp} X_{pm} Z_{jm} Z_{lm}\,,
\end{align}
where we used the condition $\epsilon_{mn}\ssp\epsilon_{nq}\ssp\epsilon_{qp}\ssp\epsilon_{pm} = +1$.  
These plaquette operators generate the stabilizer group,
\begin{align}
  \mathcal S = \big\{S_{k_0\ssp k_1\ssp k_2\ssp k_3}\;\vert\;\text{for all plaquettes}\big\}\,.
\end{align}

The syndrome of a Pauli error can be described by the set of plaquette stabilizers with which it anticommutes.  The single-qubit error $Z_{lm}$ anticommutes with two stabilizers, $S_{i\ssp j\ssp m\ssp l}$ and $S_{l\ssp m\ssp p\ssp o}$.  In general, a Pauli-$Z$ error  anticommutes with all the stabilizers involving the edge on which it acts.  The single-qubit error $X_{lm}$ anticommutes with only two stabilizers, $S_{i\ssp j\ssp m\ssp l}$ and $S_{m\ssp n\ssp q\ssp p}$.  The same holds true for the single-qubit error $X_{jm}$; therefore, these two errors have the same syndrome and can only be detected but not corrected.  The Pauli error $Y_{lm}$ anticommutes with $S_{l\ssp m\ssp p\ssp o}$ and $S_{m\ssp n\ssp q\ssp p}$, which has the same syndrome as $Z_{mp}$.  In App.~\ref{sec:bksf_infinite_code}, we list the syndromes of all the single qubit errors for the \BKSF\ on an infinite lattice with the specific edge ordering in Fig.~\ref{fig:bk_lattice_c}.  These results show that all single-qubit errors can be detected, but some of them share the same syndromes and cannot be corrected.  One has to be careful for open-boundary conditions; for example, the Pauli-$Y$ errors at the bottom of an open lattice commute with all stabilizers and cannot be even detected.

\section{Majorana loop stabilizer codes}
\label{sec:mlsc}

A natural question to ask is whether the \BKSF\ on a square lattice can correct all single-qubit errors.  The no-go thereom in~\cite{setia_superfast_2019} implies that the \BKSF\ on a 2D (or even 3D) square lattice cannot correct all single-qubit errors.  An example was also given in~\cite{setia_superfast_2019} to get around the no-go theorem lattice by introducing extra ancillary vertices and edges, which increase the degree of the original vertices from $4$ to $8$ on 2D square lattices.  Here, we introduce a geometric locality-preserving fermion-to-qubit map that correct all single-qubit errors on 2D square lattices, without introducing any additional vertices or edges.  Compared to the \BKSF, logical operations can also be implemented with lower-weight Pauli operators with our code.  In Apps.~\ref{sec:general_mlsc}, we discuss two methods to construct these codes in general.

\begin{figure}[htb]\label{fig:mlsc_loc}
\centering
\includegraphics[width=0.22\textwidth]{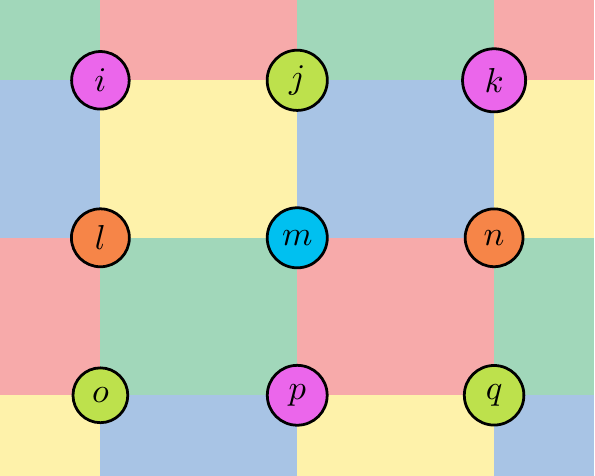}
\caption{The structure of the code: each circle represents a vertex in the 2D lattice; its color denotes one of the four types of vertex. Similarly, the color of a plaquette indicates the type of the corresponding stabilizer. }
\end{figure}

Similar to the \BKSF, the logical operators in our code satisfy the commutation relations of the quadratic Majorana operators in Eq.~\eqref{eq:commu_quadratic_majo}.   The stabilizers~\eqref{eq:closed_path_relation_qubit} of the code also correspond to the products of Majorana operators on a loop.   Therefore, we call our construction the Majorana loop stabilizer code (\MLSC); the \BKSF\ is a special case of the \MLSC.  The logical operators are allowed to be more general, e.g., non-uniform with respect to vertices and may involve Pauli-$Y$ operators.  This modification allows the \MLSC\ to evade the no-go theorem in~\cite{setia_superfast_2019}.  We discuss the code using the 2D square lattice as an example, see Fig.~\ref{fig:mlsc_loc};
each circle therein represents a vertex (fermionic mode), and its color indicates the type of the vertex.  There is one stabilizer operator associated with each plaquette (4 vertices), and its color indicates the type of the stabilizer.  

\begin{figure}[htb]\label{fig:mlsc_a_all}
\centering
\subfloat[]{\label{fig:mlsc_a}
\includegraphics[width=0.25\textwidth]{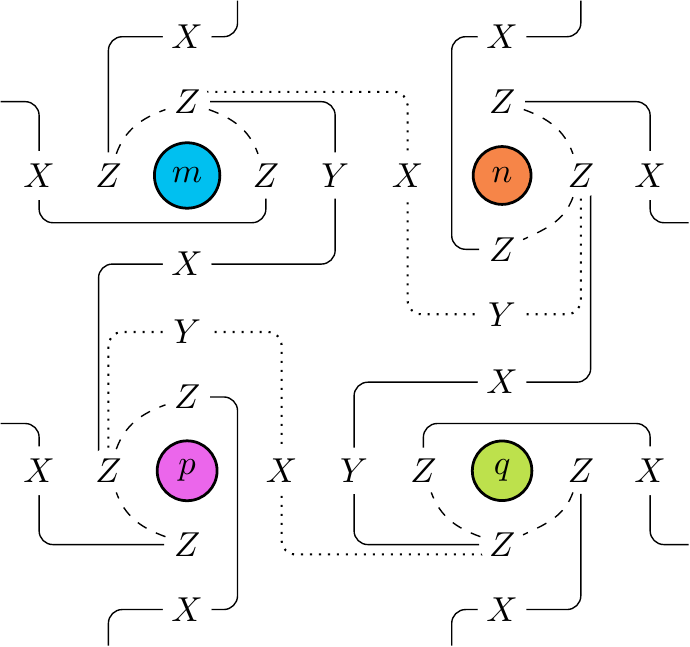}}
\hspace{0.8cm}
\subfloat[]{\label{fig:mlsc_stab_a}\raisebox{1cm}{
\includegraphics[width=0.095\textwidth]{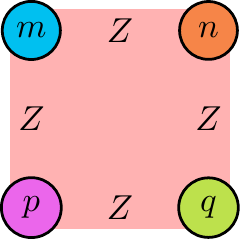}}}
\caption{\protect\subref{fig:mlsc_a} The logical operators associated with the red plaquette in Fig.~\ref{fig:mlsc_loc}.  Similar to the \BKSF, there is one physical qubit situated on each edge of the lattice, and $X$, $Y$, and $Z$ denote the corresponding Pauli operators.  Each vertex operator $\eta$ is mapped to a product of three Pauli-$Z$ operators linked by a curved dashed line.   Each edge operator $\xi$ is mapped to a product of single-qubit Pauli operators linked by a solid/dotted line, where the unique Pauli-$X$ operator acts on the corresponding edge of $\xi$.  The two rightmost (leftmost) Pauli-$X$ operators are shared by the vertices $n$ and $q$ ($m$ and $p$) and the their right (left) neighbors, and similarly for the two top (bottom) Pauli-$X$ operators.  This plaquette has a rotational symmetry of order $4$ (unchanged by a rotation of angle $\pi/2$). \protect\subref{fig:mlsc_stab_a} The stabilizer operator $S_{mnqp}$ associated with the plaquette up to a minus sign.}
\end{figure}

The logical operators and the stabilizer related to a red plaquette, e.g., the one consisting of the vertices $m$, $n$, $q$, and $p$, are plotted in Fig.~\ref{fig:mlsc_a_all}.
The Pauli operators $X$, $Y$, and $Z$ therein act on the qubits on the corresponding edges.  Each quadratic Majorana operator in Eq.~\eqref{eq:quadratic_majo_generators} is mapped to a product of single-qubit Pauli operators linked by a line.  The vertex  operator $\eta_m$ is mapped to a product of three Pauli-$Z$ operators linked by a curved dashed line around the site $m$, and similar for the other vertex operators.  The edge operator $\xi_{mn}$ is mapped to the Pauli operator $Z\otimes X\otimes Y\otimes Z$ on the upper dotted line, where the only Pauli-$X$ operator acts on the edge $(m, n)$.  The same rule applies to the remaining edge operators $\xi$, where the Pauli-$X$ operator always acts on the corresponding edge of $\xi$.  These operators satisfy the commutation relations from \eqref{eq:commu_quadratic_majo}, which is easily verified numerically.  The stabilizer associated with the plaquette is plotted in Fig.~\ref{fig:mlsc_stab_a}, up to an overall minus sign. 
\begin{align}
  S_{mnqp}  &= \tilde\xi_{mn} \,\tilde\xi_{nq}\, \tilde\xi_{qp}\, \tilde\xi_{pm}= -Z_{mn}\ssp Z_{nq}\ssp Z_{qp}\ssp Z_{pm}\,,
\end{align}
where we use the same definition~\eqref{eq:closed_path_relation_qubit} and convention as in the \BKSF\ example; $\epsilon_{jk} = 1$ if the site $k$ is to the right of, or below the site $j$.  For nomenclature purpose, we refer terms such as $\xi_{mn}$, $\xi_{mn}\eta_m$, $\xi_{mn}\eta_n$, or $\xi_{mn}\eta_m\eta_n$ as generalized edge operators.  The weight of any generalized edge operator in the red plaquette is greater or equal to three.  This is a necessary condition for correcting all single-qubit errors.  The \BKSF\ violates this condition on a 2D lattice no matter how the incident edges are ordered. 

\begin{figure}[htb]\label{fig:mlsc_b_all}
\centering
\subfloat[]{\label{fig:mlsc_b}
\includegraphics[width=0.3\textwidth]{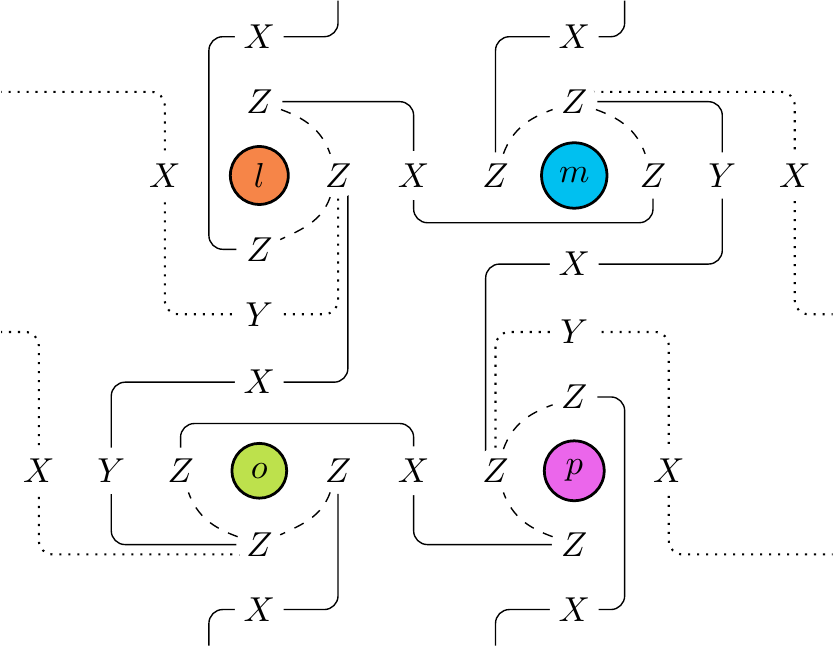}}
\hspace{0.27cm}
\subfloat[]{\label{fig:mlsc_stab_b}\raisebox{0.5cm}{
\includegraphics[width=0.14\textwidth]{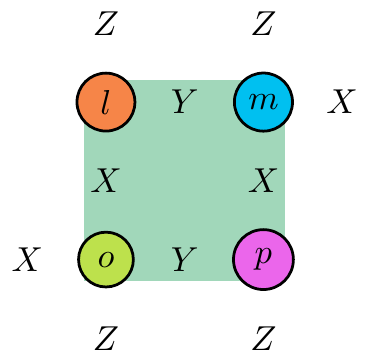}}}
\caption{\protect\subref{fig:mlsc_b} The logical operators of the green plaquette consisting of the vertices $l$, $m$, $o$, and $p$.  This plaquette has a rotational symmetry of order $2$ (unchanged by a rotation of an angle $\pi$). \protect\subref{fig:mlsc_stab_b} The stabilizer $S_{lmpo}$ of the plaquette up to a minus sign.}
\end{figure}

\begin{figure}[htb]\label{fig:mlsc_c_all}
\centering
\subfloat[]{\label{fig:mlsc_c}
\includegraphics[width=0.27\textwidth]{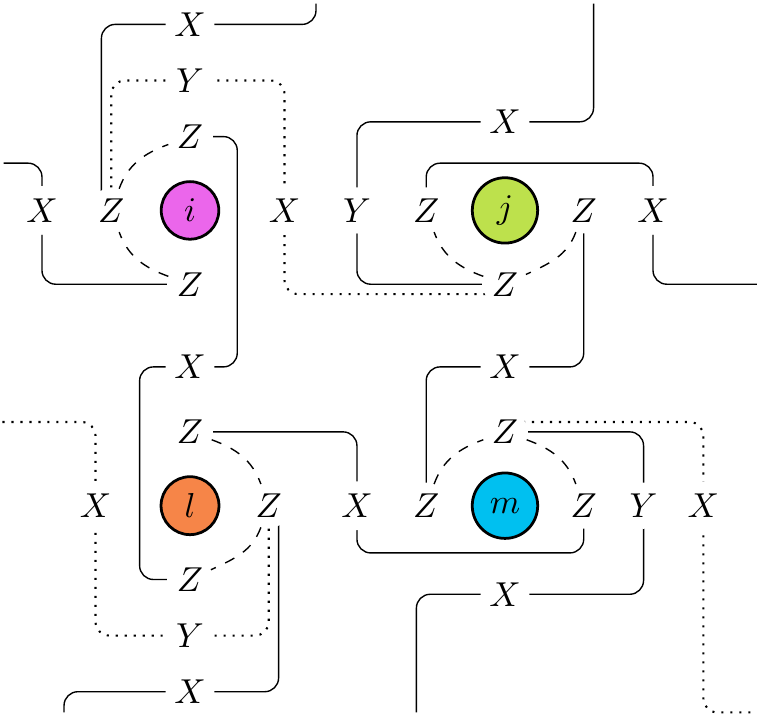}
}\hspace{0.24cm}
\subfloat[]{\label{fig:mlsc_stab_c}\raisebox{0.9cm}{
\includegraphics[width=0.14\textwidth]{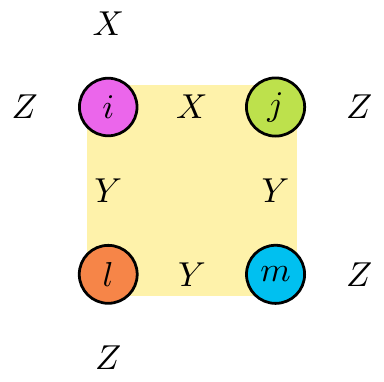}}}
\caption{\protect\subref{fig:mlsc_c} The logical operators of the yellow plaquette consisting of the vertices $i$, $j$, $l$, $m$. \protect\subref{fig:mlsc_stab_c} The corresponding stabilizer $S_{ijml}$ up to a minus sign.}
\end{figure}

\begin{figure}[htb]\label{fig:mlsc_c_bar_all}
\centering
\subfloat[]{\label{fig:mlsc_c_bar}
\includegraphics[width=0.27\textwidth]{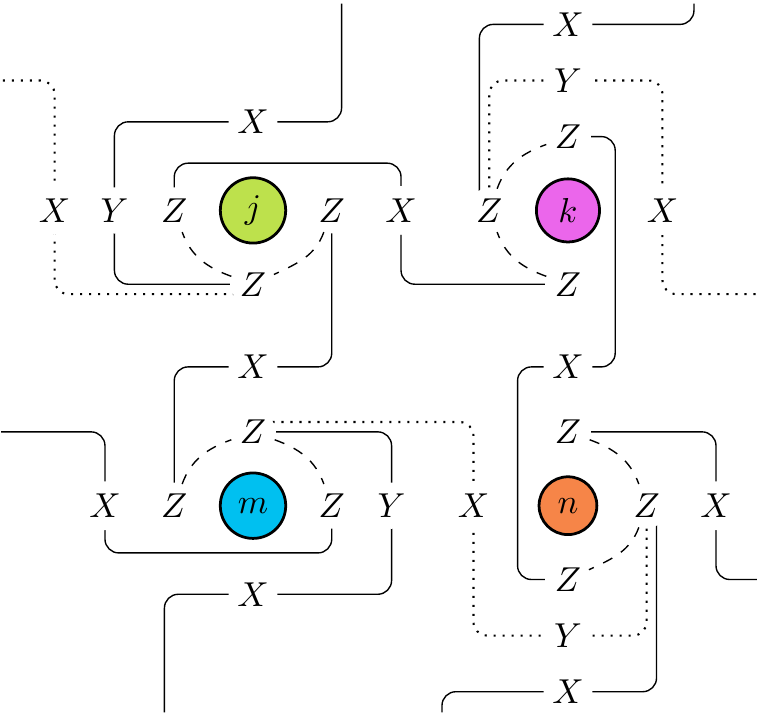}
}\hspace{0.24cm}
\subfloat[]{\label{fig:mlsc_stab_c_bar}\raisebox{0.9cm}{
\includegraphics[width=0.14\textwidth]{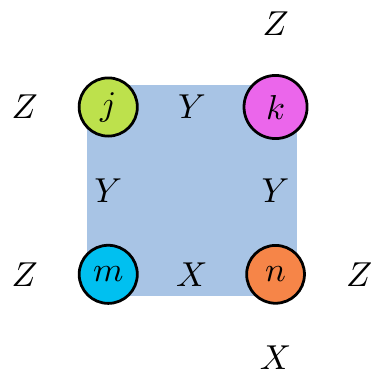}}}
\caption{\protect\subref{fig:mlsc_c_bar} The logical operators of the blue plaquette consisting of vertices $j$, $k$, $m$, and $n$.  These logical operators are identical to those in Fig.~\ref{fig:mlsc_c} by a $\pi$ rotation.  \protect\subref{fig:mlsc_stab_c_bar} The corresponding stabilizer $S_{jknm}$ up to a minus sign, which is identical to the one in Fig.~\ref{fig:mlsc_stab_c} by a $\pi$ rotation.}
\end{figure}

The entire code can be constructed by putting together the four types of vertices illustrated in Fig.~\ref{fig:mlsc_a}, where each Pauli-$X$ operator is shared by two neighboring vertices.  For clarity, we plot the logical operators in the green plaquette consisting of the vertices $l$, $m$, $o$, and $p$ in Fig.~\ref{fig:mlsc_b_all};   
the logical operators and the stabilizer of the yellow plaquette consisting of the vertices $i$, $j$, $l$, and $m$ are plotted in Fig.~\ref{fig:mlsc_c_all};
the logical operators of the blue plaquette consisting of the vertices $j$, $k$, $m$, and $n$ are plotted in Fig.~\ref{fig:mlsc_c_bar_all}.  
Any generalized edge operator in the green, yellow, or blue plaquette also has weight $w\geq 3$.  

We now go a step further to argue that any logical operator $M\not\in \mathcal S$  must have weight $w\geq 3$.  Up to an overall sign such an operator can be specified as
\begin{align}
  M = \prod_{(j,k)\in E_M} \tilde \xi_{jk} \prod_{l\in V_M} \tilde \eta_l\,,
\end{align}
where $E_M$ and $V_M$ are the sets of edges and vertices that determines $M$.  Without loss of generality, we assume that for any vertex $j\in V_M$ there must be an edge $(j,k)\in E_M$ for some vertex $k$; otherwise, the weight of $M$ can be reduced by removing $j$ from $V_M$ (the only exception is that removing $\tilde{\eta}_k$ from $\tilde{\eta}_j \tilde{\eta}_k \tilde{\eta}_m$ in Fig.~\ref{fig:mlsc_c_bar_all} increases its weight from 5 to 6).  We refer to a vertex $j$ as a single-paired vertex if exactly one of its incident edges is in $E_M$; at least one Pauli-$Z$ operator on the other three incident edges of $j$ contributes to $M$ no matter whether $j\in V_M$ or not.  If there are more than one single-paired vertex in $V_M$, the weight of $M$ must be greater than, or equal to three; this is because there must be some Pauli-$X$ or $Y$ components in $M$ to make it anticommute with the vertex operators on the single-paired vertices, besides at least two Pauli-$Z$ operators associated with the single-paired vertices.  When $V_M$ contain two edges, the only case with less-than-two single-paired vertices is two parallel edges sitting next to each other.  One can check that the weights of such logical operators are always greater than, or equal to three in Figs.~\ref{fig:mlsc_a_all}-\ref{fig:mlsc_c_bar_all}.  The specific choice of the vertex layout in Fig.~\ref{fig:mlsc_loc} is important to preventing the opposite from happening.   

\begin{figure}[htb]\label{fig:mlsc_stab_loc}
\centering
\includegraphics[width=0.22\textwidth]{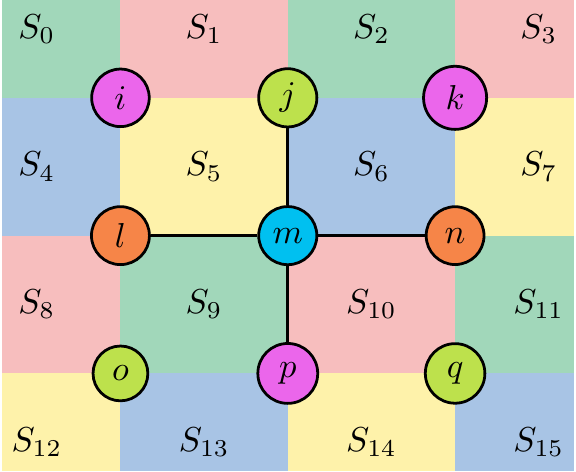}
\caption{The stabilizers on the plaquettes for syndrome detection. }
\end{figure}

To make the discussion of the syndromes of the single-qubit errors easier, we label the stabilizers in Fig.~\ref{fig:mlsc_stab_loc}.
The syndrome of a single-qubit error can be described by the stabilizer operators which anticommute with it.  Here, we list the syndromes of the single-qubit errors on all the incident edges of the vertex $m$ in Fig.~\ref{fig:mlsc_stab_loc},
%\begin{align}
%&X_{mp}:  S_{10},\, \bar S_{13}', && Z_{mp}:  S_9'', \,S_{14}'\,,\\[3pt]
%&X_{lm}:  \bar S_4', \, S_5',  \bar S_6', \,S_9'', && Z_{lm}:  S_5', \,S_9''\,,\\[3pt]
%&X_{mn}:  S_{10}, \,S_5',  &&Z_{mn}:   S_9'', \, \bar S_6'\,,\\[3pt]
%&X_{jm}:   S_2'', \,S_5', \,\bar S_6', \,S_9'', && Z_{jm}: S_5', \,\bar S_6'\,.
%\end{align}
\begin{align}
&X_{mp}:  S_{10},\, S_{13}, && Z_{mp}:  S_9, \,S_{14}\,,\\[3pt]
&X_{lm}:   S_4, \, S_5,  S_6, \,S_9, && Z_{lm}:  S_5, \,  S_9\,,\\[3pt]
&X_{mn}:  S_{5}, \,S_{10},  &&Z_{mn}:   S_6, \,  S_9\,,\\[3pt]
&X_{jm}:   S_2, \,S_5, \, S_6, \,S_9, && Z_{jm}: S_5, \, S_6\,.
\end{align}
These single-qubit errors (Pauli-$Y$ errors not listed) have different syndromes.  We have checked that all the single-qubit errors have different syndromes on an $8\times 8$ lattice with periodic boundary conditions (128 qubits). 

To simulate lattice fermion problems, we often need to implement the hopping term $\cc_j^\dagger\cc_k + \cc_k^\dagger\cc_j$ and the occupation term $\cc_k^\dagger\cc_k$, see Eqs.~\eqref{eq:occupation} and \eqref{eq:hopping}.  With the 2D \BKSF\ in Fig.~\ref{fig:bk_lattice_c}, both the vertical and horizontal hopping terms are mapped to qubit operators of weight $6$.  With our code, the hopping terms are mapped to qubit operators of weight no more than $4$ (sometimes $3$); the occupation terms can also be implemented with lower-weight qubit operators.  Being able to implement logical operations with low-weight Pauli operators while correcting all single-qubit errors is a desired feature for near-term quantum error-mitigating schemes.  The weights of the stabilizers in our code range from $4$ to $10$.  Measurements are often the bottleneck in error-corrected quantum computation implemented on near-term devices.  For example, it can take up to $\sim 1\si{\micro\second}$ to readout the state of a superconducting qubit with fidelity $\sim 0.95$, while the typical two-qubit gate time is $\sim 40\si{\nano\second}$ with fidelity $\sim 0.995$.  Syndrome readouts should only be performed sparsely on these near-term devices, e.g, per syndrome readout cycle for at least tens of logical gate cycles.  The total number of gates, therefore, is dominated by the logical operations and not the syndrome readout.  In a recent work by some of the authors, a scheme is proposed to decipher syndromes using classical post-processing~\cite{mcclean_decoding_2020}.  This scheme avoids both complicated syndrome measurements and additional ancilla qubits for the measurements, and can be used to readout the higher-weight stabilizers. 

In Fig.~\ref{fig:comparison}, we summarize the differences between our code and the \BKSF\ on a 2D lattice. 
\begin{figure}[ht]
\begin{center}
{\renewcommand{\arraystretch}{1}%
\begin{tabular}{c | c | c | c | c}
\hline\hline
   &{\hspace{1mm}} distance{\hspace{1mm}} & {\hspace{1mm}}occupation{\hspace{1mm}}  &{\hspace{1mm}} hopping{\hspace{1mm}} & stabilizer  \\
  \hline		
  \BKSF\ & 2 & 4 & 6 & 6 \\
  \hline
  Our code & 3 & 3 & 4 (3) & 4--10   \\
  \hline
\end{tabular}
}
\end{center}
\caption{\label{fig:comparison}A comparison of our code to \BKSF\ on a 2D square lattice in terms of the code distance, weights of the occupation terms, and weights of the hopping terms.  The code here exhibits improved distance and weight (logical) properties.  The weights of the stabilizers in our code, ranging from $4$ to $10$, are typically higher than BKSF.}
\end{figure}

\section{Open-boundary conditions}
\label{sec:obc}

In the last section, we discussed MLSCs defined on lattices with periodic boundary conditions.  In realistic experimental setups, however, periodic boundary conditions are usually unavailable or very costly to have.  In this section, we provide a procedure to construct MLSCs with open boundary conditions that correct all single-qubit errors.  

\begin{figure}[htb]\label{fig:mlsc_stab_loc_open_4by4}
\centering
\includegraphics[width=0.32\textwidth]{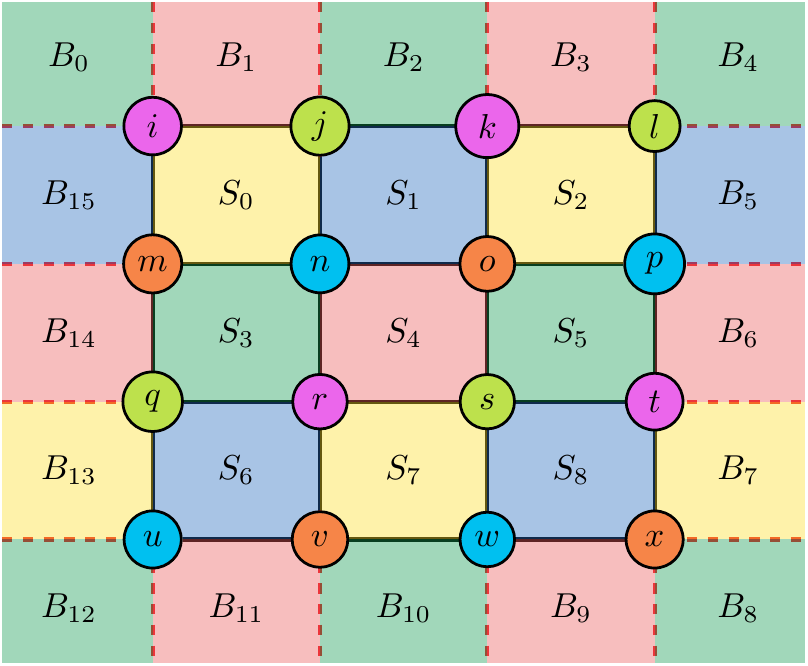}
\caption{MLSC with open boundary conditions: each vertex represents one fermionic mode (16 in total), each edge (solid black) in the bulk of the code is associated with a logical edge operator (24 in total), and each added dangling edge (dashed red) introduces an additional physical qubit at the boundaries (16 in total).}
\end{figure}

Boundary effects are significant for lattices of modest sizes; for instance, most of the vertices are located at the boundaries in a $4\times 4$ square lattice.  Errors on the boundary qubits are more difficult to correct, because they are involved in a smaller number of stabilizers, i.e., less syndromes can be used to correct/detect the errors.  This problem is especially prominent for qubits around the corners of a lattice, where two boundary edges meet.  For this reason, using the MLSC introduced in the last section on lattices with open boundary conditions results in uncorrectable single-qubit errors.  Here, we discuss how to circumvent that using an example where dangling boundary edges are added to a $4\times 4$ square lattice, see Fig.~\ref{fig:mlsc_stab_loc_open_4by4}.  
Just like the bulk of the code, we put one physical qubit on each dangling boundary edge (dashed red).  The logical operators, including all the vertex operators and edge (black solid) operators, are defined using the same rules as the MLSC introduced in the last section.  This is possible because any vertex (edge) operator of a MLSC only involves the Pauli operators on the edges incident on that vertex (edge), i.e,  logical operators in the bulk of the code has no support beyond the dangling edges.  Therefore, the commutation relations~\eqref{eq:commu_quadratic_majo} are satisfied, and the stabilizers in the bulk of the code $S_0\,, S_1\,,\ldots, S_{8}$ remain the same as those in the periodic boundary case.  Introducing the dangling edges adds extra degrees of freedom, which allow us to construct additional stabilizers $B_0\,, B_1\,,\ldots, B_{15}$ to correct errors at the boundaries.  Since the number of boundary stabilizers is equal to the number of dangling edges, the extra degrees of freedom are eliminated.  Because each qubit at (or close to) the boundary is involved in more stabilizers, errors on them are more likely to be correctable.  

\begin{figure}[htb]\label{fig:mlsc_dangling_edge_j}
\centering
\includegraphics[width=0.1\textwidth]{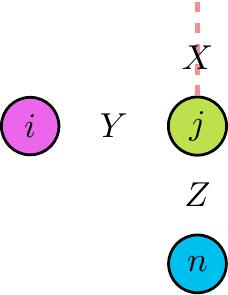}
\caption{The edge operator associated with the dangling edge (dashed red) incident on vertex $j$.}
\end{figure}

To define the boundary stabilizers, we first introduce the edge operators associated with the dangling edges.  We denote these operators as $D_0\,, D_1\,,\ldots, D_{15}$ starting with the vertical dangling edge of the vertex $i$ and counting clockwise along the boundaries.  These operators are defined using a similar rule as the edge operators in the bulk of the code, but Pauli operators on edges beyond the dangling edges are neglected.  For example, the operator $D_0$ associated with the vertical dangling edge incident on vertex $j$ is plotted in Fig.~\ref{fig:mlsc_dangling_edge_j}.  
Each dangling edge operator anticommutes with the vertex and edge operators incident on it.  As a consequence, the dangling edge operators commute with all the stabilizers in the bulk of the code $S_0\,, S_1\,,\ldots, S_{8}$.  

We then introduce the operators $A_0\,, A_1\,,\ldots, A_{15}$, defined as the products of all the edge operators in a boundary plaquette.  For example, $A_0$ is the product of the two dangling edge operators incident on vertex $i$ ($D_0$ and $D_{15}$), while $A_1$ is the product of the edge operators on $(i,\, j)$ and the two vertical dangling edge operators incident on vertices $i$ and $j$ ($D_0$ and $D_1$).  These operators commute with all the logical operators, because each logical operator anticommutes with either zero or two edge operators in $A_\alpha$.  However, neighboring operators $A_\alpha$ and $A_{\alpha\pm 1}$ (mod 16) anticommute with each other.  For example, $D_0$ anticommutes with $A_1$, while $D_{15}$ commute with $A_1$; therefore, $A_0 = D_0 D_{15}$ anticommutes with $A_1$. 

To construct boundary stabilizers that commute with each other, we introduce the conjugate operator for each dangling edge; these are denoted as $C_0\,, C_1\,,\ldots, C_{15}$, starting with the vertical dangling edge of the vertex $i$ and counting clockwise along the boundary.  The conjugate operator $C_\alpha$ anticommutes with $D_\alpha$, but commutes with all the logical operators and $D_\beta$ for $\beta\neq \alpha$.  The conjugate operators also commute with each other $[C_\alpha, \, C_\beta]=0$ for all $\alpha, \beta = 0,\ldots, 15$.
When all of the edge operators incident on a dangling edge only have the Pauli-$Z$ component on it, the conjugate operator is simply the single-qubit Pauli-$Z$ operator on the corresponding dangling edge.  For example, $C_{15}$ is equal to the Pauli-$Z$ operator on the horizontal dangling edge incident on vertex $i$.  The situation is more complicated when some edge operator has a Pauli-$Y$ component on the dangling edge.  For example, $C_0$ is the product of the Pauli-$Z$ operators on the edge $(i,\, j)$ and the two vertical dangling edges incident on $i$ and $j$.  In general, these operators are defined as the product of Pauli-$Z$ operators on the dangling edge, a second edge whose edge operator has a Pauli-$Y$ component on the dangling edge, and a neighboring dangling edge (if it exists) whose edge operator has a Pauli-$Y$ component on the second edge.  

\begin{figure}[htb]\label{fig:mlsc_stab_b0}
\centering
\includegraphics[width=0.15\textwidth]{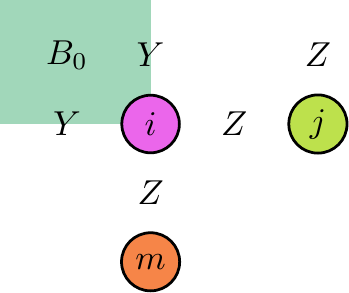}
\caption{The stabilizer $B_0$ associated with the green plaquette on the upper-left corner of the lattice. }
\end{figure}

\begin{figure}[htb]\label{fig:mlsc_stab_b1}
\centering
\includegraphics[width=0.1\textwidth]{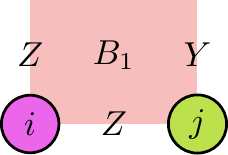}
\caption{The stabilizer $B_1$ associated with the red plaquette on the top of the lattice. }
\end{figure}

\begin{figure}[htb]\label{fig:mlsc_stab_b2}
\centering
\includegraphics[width=0.25\textwidth]{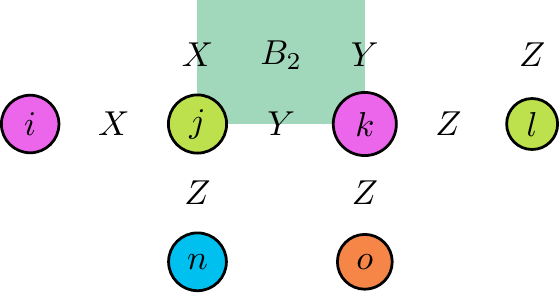}
\caption{The stabilizer $B_2$ associated with the green plaquette on the top of the lattice. }
\end{figure}

We now define the boundary stabilizers as
\begin{align}
    B_\alpha = i A_\alpha C_\alpha\,,\quad \text{for $\alpha=0, 1,\ldots,15$}.
\end{align}
These operators commute with all the logical operators and the stabilizers in the bulk.  They also commute with each other; e.g., we have $[B_\alpha,\, B_{\alpha+1}] =0$ using the commutation relations
\begin{gather}
   \{A_\alpha,\, A_{\alpha+1}\} = 0\,,\quad \{C_\alpha,\, A_{\alpha+1}\} = 0\,,\\[3pt]
   [A_\alpha,\, C_{\alpha+1}] = 0\,,\quad [C_\alpha,\, C_{\alpha+1}] = 0\,.
\end{gather}
In Figs.~\ref{fig:mlsc_stab_b0}-\ref{fig:mlsc_stab_b2}, we plot the boundary stabilizers $B_0$, $B_1$, and $B_2$.
We have checked numerically that any Pauli operator with weight one or two anticommutes with at least one of the stabilizers; therefore, the code distance is three and all single-qubit errors are correctable.  

The types of corner vertices have to be chosen very carefully: the upper-left vertex has to be red, the upper-right vertex has to be green\ldots  Any choice of corner vertices different from those in Fig.~\ref{fig:mlsc_stab_loc_open_4by4} leads to single-qubit errors with colliding syndromes.  One can construct open boundary MLSCs of larger sizes using the same procedure discussed here for the $4\times 4$ lattice.

\section{Encoding a Slater determinant}
\label{sec:encoding}

In this section, we discuss how to encode a single Slater determinant with \MLSC\ in circuit depth $O(1)$.  For simplicity, we only consider Slater determinants in the computational basis; arbitrary Slater determinants can be prepared by applying single-particle basis transformations to these determinants. The encoded Slater determinant is a common eigenstate of the stabilizers and the vertex operators,
\begin{gather}
     S_\ell\ket{\psi_\mathrm{det}} = \ket{\psi_\mathrm{det}},\; \text{for all}\; \ell\in L\,,\\[3pt]
    \tilde\eta_k \ket{\psi_\mathrm{det}} = \zz_k \ket{\psi_\mathrm{det}},\; \text{for all}\; k\in V\,,
\end{gather}
where $L$ consists of a complete set of independent closed paths in $G$, and $\zz_k = \pm 1$ indicates whether the $k$th fermionic site is unoccupied or occupied.  We start from a product state in the computational basis
\begin{align}
 Z_{jk} \ket{\psi_\mathrm{prod}} = z_{jk} \ket{\psi_\mathrm{prod}}\,,
\end{align}
where $Z_{jk}$ is the Pauli-$Z$ operator on the edge $(j, k)$.  

We first consider the case of \BKSF.  The eigenvalues $z_{jk} = z_{kj}$ are chosen such that
\begin{align}\label{eq:vertex_parity}
  \prod_{j:\,(j,k)\in E} z_{jk} = \zz_k\,.
\end{align}
There are some freedoms in assigning the values $z_{jk}$; one for each independent closed paths in $G$.  The constraints in Eq.~\ref{eq:vertex_parity} can be converted to a system of linear equations over $GF(2)$ by mapping $z_{jk}\pm 1$ to the binary values $0$ and $1$.  When a Hamiltonian path $(0,1,\ldots, \length-1)$ of the graph $G$ is known, we can assign the values of $z_{jk}$ in the Hamiltonian path according to
\begin{align}
 z_{0,1} = z_0\,, \quad z_{k, k+1} = z_{k}\, z_{k-1, k}\,,
\end{align}
where $k=1,2,\ldots, \length-2$.  We then assign the value  $+1$ to all of the remaining $z_{jk}$, and the condition~\eqref{eq:vertex_parity} is satisfied.  More generally, we can assign values to $z_{jk}$ with the following three steps.
\begin{enumerate}
     \item Find a spanning tree $T$ of the graph $G$, which takes time linear in $\norm{V}$.
     \item Assign the values in $T$ starting from the leaves towards the root; if $j$ is a parent node of $k$, let $z_{jk}$ be the product of $z_k$ and $z_{kl}$ for all the children $l$ of $k$.
     \item Set the value of the remaining edges that are not in $T$ to $+1$.
\end{enumerate}
This algorithm requires that the total parity of the vertex operators be even as each Pauli-$Z$ on the edges of the spanning tree appears in two vertex operators.  In App.~\ref{sec:initial_prep}, we also present an algorithm to assign values to $z_{jk}$ without using a Hamiltonian path or a spanning tree. 

With the \MLSC\ in Sec.~\ref{sec:mlsc}, not all the incident edges participate in the vertex operator $\tilde \eta_k$, and the condition~\eqref{eq:vertex_parity} is modified to 
\begin{align}\label{eq:vertex_parity_mlsc}
  \prod_{j:\,(j,k)\in E_k} z_{jk} = \zz_k\,,\quad\text{for all $k\in V$}\,,
\end{align}
where $E_k$ is the set of edges that are involved in $\tilde \eta_k$.  We use the notion $E'$ to denote the set of edges $(j,k)$ such that $(j,k) \in E_j$ and $(j,k) \in E_k$.  One can assign values to $z_{jk}$ with the following four steps.
\begin{enumerate}
\item Check the total parity of the vertex operators for each connected part of the graph $(V, E')$.
\item If the parity is odd, find an edge $(j,k)$ in the connected part such that $(j,k)$ is involved in $\tilde\eta_j$ but not $\tilde\eta_k$, set the value of $z_{jk}$ to $-1$.
\item Find a spanning tree of the connected part and use the algorithm for \BKSF\ to assign the values to the edges in the spanning tree,  
\item after going through all the connected parts, set the values of the remaining edges to $+1$. 
\end{enumerate}
The resulting state $\ket{\psi_\mathrm{prod}}$ is an eigenstate of all the vertex operators with the prescribed eigenvalues,
\begin{align}
    \tilde \eta_k\ket{\psi_\mathrm{prod}} = z_k\,,\quad \text{for all $k\in V$}\,.
\end{align}
We then measure the values of the stabilizer operators $S_\ell$ and record the measurement results as $s_\ell$.   Since the stabilizers commute with the operators $\tilde \eta_j$, the post-measurement state satisfies
\begin{gather}
 S_\ell\ssp\ket{\psi_\mathrm{post}} = s_\ell\ssp\ket{\psi_\mathrm{post}}\,,\quad \text{for all $\ell\in L$}\,,\\[2pt]
 \tilde \eta_j\ssp\ket{\psi_\mathrm{post}} = z_j\ssp\ket{\psi_\mathrm{post}}\,,\quad \text{for all $j\in V$}\,.
\end{gather}

Generally, some of the measurement results of the stabilizers will be $-1$.  One simple way to resolve this problem is to reassign the values $\epsilon_{jk}$ in Eq.~\eqref{eq:xi_qubit}, and the signs of the stabilizers change accordingly.  This procedure corresponds to encoding the fermionic state in a different subspace of the stabilizers; the signs of logical operators change accordingly.  The reassignment is straightforward when there is a known Hamiltonian path in $G$.  We keep  $\epsilon_{jk}$ unchanged for all the edges $(j,k)$ in the Hamiltonian path.  Any remaining edge $\epsilon_{lm}$ along with a section in the Hamiltonian path form an independent closed path.  We flip the value of $\epsilon_{lm}$ if the measurement outcome of the corresponding stabilizer is $-1$.  More generally, the reassignment of $\epsilon_{jk}$ can be achieved with three final steps.
\begin{enumerate}
     \item Find a spanning tree $T$ of the graph $G$.
     \item Assign the original value $\epsilon_{jk}$ to all of the edges $(j,k)$ in the spanning tree $T$.
     \item Flip the sign of any remaining term if the measurement outcome of the stabilizer is $-1$. 
\end{enumerate} 

\section{Extension to more general fermion models}
\label{sec:chemistry}

In a more general context than the lattice models discussed thus far, one may wish to consider the quantum simulation of interacting fermion models with a high degree of connectivity. For example, the molecular electronic structure Hamiltonian, central to quantum chemistry, can be expressed as
\begin{equation}
H = \sum_{pq} h_{pq} a^\dagger_p a_q + \sum_{pqrs} h_{pqrs} a^\dagger_p a^\dagger_q a_r a_s
\label{eq:chem}
\end{equation}
where the scalars $h_{pq}$ and $h_{pqrs}$ are obtained by integrating the Coulomb and kinetic operators over a discrete basis set \cite{Helgaker2002}. This Hamiltonian has ${ O}(N^2)$ terms in the one-body operator where $N$ is the number of orbitals (basis functions) into which the system is discretized. Thus, directly applying the MLSC or BKSF to~\eqref{eq:chem} would require $O(N^2)$ qubits, which is an unacceptable overhead in most contexts.

However, rather than directly encoding the Hamiltonian for the system we wish to simulate, we might instead consider encoding the algorithm that we would use to perform the simulation. Specifically, if we can realize a simulation of~\eqref{eq:chem} (or more generally, any algorithm) using only a gate set defined in terms of evolution under fermionic operators on a lattice, then we can encode that algorithm into the MLSC or BKSF representation with only a constant factor increase in the number of qubits, thus making the error-correction properties of these encodings practical to utilize in a much more general context. In fact, some of the most efficient algorithms for simulating the Hamiltonian of~\eqref{eq:chem} already meet these requirements. We will now discuss one such algorithm.

Without loss of generality one can perform a tensor factorization of the Coulomb operator to express the Hamiltonian $H$ of~\eqref{eq:chem} as
\begin{equation}
R_0 \!\left(\sum_{p} f_p n_p\right)\! R_0^\dagger + \sum_{\ell=1}^{L} R_\ell\! \left(\sum_{pq} g_{pq}^{(\ell)} n_p n_q\right)\! R_\ell^\dagger
\end{equation}
where $f_p$ and $g_{pq}^{(\ell)}$ are scalars, $L = O(N)$ for chemistry but $L = N^2$ in general, and $R_\ell$ is a unitary rotation of the single-particle orbital basis. The work of \cite{Motta2018} suggests simulating the chemistry Hamiltonian in this representation by performing a Trotter step of time-evolution $e^{-i \delta t H}$ for short duration $\delta t$ as
\begin{equation}
R_0\! \left(e^{-i\delta t \sum_{p} f_p n_p}\right)\! R_0^\dagger \prod_{\ell} R_\ell \!\left( e^{-i\delta t \sum_{pq} g_{pq}^{(\ell)} n_p n_q} \right)\! R_{\ell}^\dagger.
\end{equation}

Key to this approach are algorithms introduced in \cite{kivlichan_quantum_2018} which enable exact realization of both $R_\ell$ and time-evolution under $\sum_{pq}g_{pq}n_p n_q$, each using $O(N^2)$ applications of nearest-neighbor gates from a set generated only by the following nearest-neighbor fermion terms: $n_p$, $n_{p+1}$, $n_p n_{p+1}$, and $(a^\dagger_p a_{p+1} + a^\dagger_{p+1}a_p)$. Crucially, all of these fermion terms are represented within the BKSF or MLSC applied to a planar lattice. Thus, the entire Trotter step requires only ${O}(L N^2)$ gates within either the BKSF or MLSC (which translates to ${O}(N^3)$ gates in the case of chemistry, matching the lowest known gate complexity of a Trotter step of this Hamiltonian~\cite{Motta2018}) and requires only $2N$ qubits. One would use the BKSF or MLSC encoding on a 2D lattice and then snake a 1D path through this lattice over which these nearest-neighbor algorithms are implemented in the code.

While physical fermionic particles only exhibit one- and two-body interactions (and thus can always be simulated with the approach mentioned above), certain effective models may involve 3-body or higher fermion interactions. These models can also be efficiently simulated within the MLSC or BKSF representation with only $2N$ qubits. For this, one can use the higher order fermionic swap network strategies introduced recently in~\cite{ogorman_generalized_2019} which generalize the results of ~\cite{kivlichan_quantum_2018} to enable Trotter steps of time-evolution under arbitrary $k$-body fermion models using a number of nearest-neighbor gates scaling as ${O}(N^{2k})$. In this approach, the total number of gates required for quantum chemistry ($k=2$) would be ${O}(N^4)$ which is less efficient than the method described above.

Esssentially, by encoding a lattice fermion Hamiltonian whose individual terms generate all of the gates in these algorithms we are able to leverage the error-correction properties of the MLSC and BKSF in a more general context than lattice models. While it seems reasonable that leveraging these error-correction properties could improve prospects for simulating high degree fermion models such as quantum chemistry in the NISQ era, the practical performance of these codes is still an open question, which should be analyzed in more detail and ultimately, tested experimentally.

\section{Conclusion}
\label{sec:conclusion}

While hardware that supports full fledged fault-tolerant schemes is still unavailable, error mitigation schemes are likely to be an essential part of quantum algorithms in the NISQ era~\cite{preskill_quantum_2018}.  A merit of these near-term quantum error-mitigating schemes is to be flexible to the hardware and the problem being implemented.  We consider error correction/detection in quantum simulation of many-body fermonic systems with geometrically local interactions (short range), e.g., the Hubbard model.  The geometric locality of the fermion Hamiltonian is often lost when they are mapped to qubits using conventional methods, such as the Jordan-Wigner transformation.  Locality-preserving fermion-to-qubit maps prevent this by encoding the original fermions into a subspace of the qubit operators corresponding to the common $+1$ eigenstates of a set of stabilizer operators.  This makes sure that parity is conserved when fermions transit around closed paths.  

Geometric locality-preserving mappings also allow for quantum error correction/detection, and one of the leading candidates is the superfast encoding presented by Bravyi and Kitaev~\cite{bravyi_fermionic_2002}.  Indeed, we demonstrate that the Bravyi and Kitaev encoding can detect all single-qubit errors on a 2D square lattice when the orders of the incident edges are chosen properly.  Very recently, it was proved that the Bravyi-Kitaev superfast encoding cannot correct all single-qubit errors on graphs with vertex degree less or equal to six~\cite{setia_superfast_2019}.  We get around this no-go theorem by introducing a mapping that is more general than the original Bravyi-Kitaev superfast encoding.  We call it a  Majorana loop stabilizer code, meaning that the stabilizers are constructed from products of Majorana fermion operators on closed paths.  We show that these codes can correct all single-qubit errors on 2D square lattices with vertex degree four.  The logical operators in our code can also be implemented with Pauli-operators with lower weights compared to the Bravyi-Kitaev superfast encoding.  We also discuss how to encode a single Slater determinant with the Majorana loop stabilizer code in circuit depth $O(1)$. Finally, we were able to show that these codes are actually useful for performing quantum chemistry simulations without any additional asymptotic overhead and that the error-correcting properties of these codes can be realized for any algorithm that can be realized in terms of fermionic operations on a lattice.

Future work will include generalizing the code to other lattices, designing fast gate sequences to implement the logical operators, simplifying syndrome measurements, and making the code more robust to errors.

\section*{Acknowledgments}

The authors would like to acknowledge the enlightening and helpful discussions with Dave Bacon, Sergio Boixo, Yu-An Chen, William Huggins, Eleanor Rieffel, Vadim Smelyanskiy, and Kevin Sung.  ZJ would also like to extend the deepest gratitude to Ben Reichardt, whose comments and examples were extremely helpful in constructing MLSC codes for open boundary conditions. 

\appendix

\section{A brief review of quantum error-correcting codes}
\label{sec:error-correcting}

Quantum devices are vulnerable to decoherence due to interactions with environments.  It was once believed that decoherence sets an upper limit on time and size for quantum computation~\cite{unruh_maintaining_1995}.  Such a limitation was later overcome, at least in theory, with the invention of quantum error-correcting codes~\cite{shor_scheme_1995,steane_error_1996,calderbank_good_1996}.  It was shown that quantum computation can be made robust against errors when the error rate is smaller than a constant threshold~\cite{kitaev_quantum_1997,aharonov_fault_1997}.  The surface code approach is a natural choice for fault-tolerant quantum computation, which requires only nearest-neighbor couplings and modest gate fidelity~\cite{Fowler_Mariantoni_Martinis_Cleland_2012, barends_superconducting_2014}.  
%Loss of coherence degrade quantum states to statistical mixtures of product states, which can be efficiently simulated on classical computers.
%Post selection of states with particular number of excitations is crucial in demonstrating quantum supremacy~\cite{neill_blueprint_2018}.

The error threshold of fault-tolerant protocols are derived by bounding an operator norm such as the diamond norm~\cite{aliferis_quantum_2006}, which translates to a stringent requirement when the amplitudes of coherent errors add together.  For example, if one gate has over-rotation $\theta$, then a sequence of $N$ gates could produce an over-rotation of $N\theta$ and an error probability proportional to $N^2$. In order to get this worst case behavior, however, the errors have to add together in a coherent way.  This is unlikely to happen in practice, therefore, coherent errors may be no worse for fault-tolerance than incoherent ones. Indeed, Pauli- or Clifford-twirling may be used to convert any noise channel into a simple mixture of Pauli errors or depolarizing noise~\cite{wallman_noise_2016,li_efficient_2017}.

% A local stabilizer code is one in which all stabilizer generators act only on a few nearby qubits. One strategy for combating quantum errors from local noises is to encode quantum information into correlations across multiple physical qubits. 
A quantum \code{n}{k}{d} code specifies a $2^k$-dimensional subspace of $n$ physical qubits, with which $k$ logical qubits can be encoded.  The distance $d$ of a quantum error-correcting code is the minimum weight of a Pauli error by which an element of the code space can be transformed into an orthogonal element of the code space, i.e., a logical error happens. The weight of a Pauli error is the number of qubits on which a non-identity Pauli transformation acts.  An \code{n}{k}{d}-quantum code can detect up to weight $d-1$ errors, and can correct errors of weight up to $t$ satisfying $2t < d$.

A large class of quantum error-correcting codes can be described by the stabilizer formalism.  Consider the Pauli group ${\cal P}_n$ of an $n$ qubit system, and let $\cal S$ be its subgroup generated by $r$ commuting, independent generators (the stabilizers).  The common $+1$ eigensubspace of all the elements in $\cal S$ defines the code subspace of dimension $2^k$ where $k = n-r$.  Let $\cal C(\cal S)$ be the centralizer of $\cal S$, the set of elements in ${\cal P}_n$ that commute with all elements of $\cal S$.  The logical operators can be specified by using the elements in $\cal C(\cal S)$ that satisfy the commutation relations of Pauli operators. 

\section{The auxiliary fermion approach}
\label{sec:auxiliary_fermion}

In this appendix, we take a different approach to construct locality-preserving fermion-to-qubit mappings that can also detect/correct single-qubit errors.  This approach is based on the auxiliary fermion method introduced by Ball~\cite{ball_fermions_2005}, and by Verstraete and Cirac~\cite{verstraete_mapping_2005}.  We show that the map itself is not particularly good for detecting/correcting errors.  In the next appendix, we improve this by combining two unit cells as a block. 

Consider fermions hopping between neighboring sites on a 2D lattice; one auxiliary fermionic mode is introduced for each data fermionic mode.  Each vertical hopping term is accompanied with a quadratic Majorana operator on the two corresponding auxiliary modes.  We then map the fermions to qubits using the Jordan-Wigner transformation with row-major order, snake shape, and data modes interleaved with the corresponding auxiliary modes.  The parity terms from the quadratic Majorana operator cancel the ones from the vertical hopping terms, and this makes the resulting Hamiltonian local.  To reproduce the original physics, the state of the auxiliary subsystem is restricted to the common $+1$ eigenspace of the quadratic Majorana operators.

We use $\ff_{2k}$ and $\ff_{2k+1}$ ($\gamma_{2k}$ and $\gamma_{2k+1}$) to denote the two Majorana operators associated with the $k$th data (auxiliary) mode.   Using the JW transformation, the Pauli operators of the data qubits can be expressed as
\begin{align}
 X_k = \parity_0^{k-1}\ff_{2k}\,,\;\; Y_k =\parity_0^{k-1} \ff_{2k+1}\,,\;\; Z_k = i\ff_{2k}\ff_{2k+1}\,,
\end{align}
where the parity operator is the product of all the Majorana operators in a particular interval,
\begin{align}\label{eq:parity_op}
 \parity^{k}_{j} = \prod_{m=j}^{k} \big(i\ff_{2m}\ff_{2m+1}\big)\big(i\gamma_{2m}\gamma_{2m+1}\big)\,,
\end{align}
The parity operator anticommutes with any single-mode Majorana operator in its support. Similarly, the Pauli operators of the auxiliary qubits take the form
\begin{align}
 \widetilde X_k = \parity_0^{k-1}Z_k\gamma_{2k}\,,\;\; \widetilde Y_k = \parity_0^{k-1} Z_k \gamma_{2k+1}\,,\;\; \widetilde Z_k = i\gamma_{2k}\gamma_{2k+1}\,.
\end{align}
The horizontal hopping term can be implemented with local qubit operators,
\begin{align}
 i\ff_{2j+1}\ssp\ff_{2j+2} = i \parity_{0}^{j-1} Y_j  \parity_{0}^{j} X_{j+1}  = -X_{j}\widetilde Z_j X_{j+1} \,.
\end{align}
The vertical fermionic hopping term takes the form
\begin{align}\label{eq:hopping_bvc}
 i\ff_{2j+1}\ssp\ff_{2k} = i \parity_{0}^{j-1}Y_j  \parity_{0}^{k-1} X_k  = iY_j\parity_{j}^{k-1}X_k \,,
\end{align}
where $j$ and $k$ are neighboring qubits in the same column.  To get rid of the non-local parity operator in Eq.~\eqref{eq:hopping_bvc}, we introduce the gauge operator of the auxiliary modes,
\begin{align}
 i\ssp\gamma_{2j+1}\ssp\gamma_{2k} 
 &= i \big(\parity_{0}^{j-1} Z_j\widetilde Y_j\big)\big( \parity_{0}^{k-1} Z_k\widetilde X_k\big)\nonumber \\
 &= i  Z_j \widetilde Y_j\parity_{j}^{k-1} Z_k\widetilde X_k  \,, \label{eq:gauge_ops}
\end{align}
where $j$ and $k$ are neighboring sites in the same column.  The hopping term can be implemented locally by attaching the gauge operator to it,
\begin{align}
 (i\ff_{2j+1}\ssp\ff_{2k})  (i\ssp\gamma_{2j+1}\ssp\gamma_{2k})&= X_j\widetilde Y_j Y_k \widetilde X_k \,.\end{align}
To retain the same physics, we restrict the state to be a common $+1$ eigenstate of the gauge operators,
\begin{align}
\bra{\psi} i\ssp\gamma_{2j+1}\ssp\gamma_{2k}\ket{\psi} = 1\,.
\end{align}
This constraint can be made local by combining two neighboring gauge operators together,
\begin{align}
&\hspace{-.5cm} (i\ssp\gamma_{2j+1}\ssp\gamma_{2k} )( i\ssp\gamma_{2j+3}\ssp\gamma_{2k-2} ) \nonumber\\
& = (i  Z_j \widetilde Y_j\parity_{j}^{k-1} Z_k\widetilde X_k )(i  Z_{j+1} \widetilde Y_{j+1}\parity_{j+1}^{k-2} Z_{k-1}\widetilde X_{k-1} )\nonumber\\
% &= Z_j \widetilde Y_j Z_{j+1}\widetilde Y_{j+1}Z_k\widetilde X_k Z_j\widetilde Z_j Z_{k-1}\widetilde Z_{k-1} Z_{k-1}\widetilde X_{k-1}\nonumber\\
&= -\widetilde X_j  Z_{j+1}\widetilde Y_{j+1}Z_k\widetilde X_k  \widetilde Y_{k-1}\,.
\end{align}

The introduction of the stabilizer operators allows for correcting/detecting certain kinds of errors.  The syndrome of a Pauli error can be described by the set of gauge operators that anticommute with it.  Any Majorana operator anticommutes with a gauge operator if and only if it contains exactly one of the two single-mode Majorana operators in the gauge operator.  The code cannot detect Pauli-$Z$ errors on the data qubits as they are logical operations and commute with all gauge operators.  It can detect any single-qubit Pauli-$X$ or -$Y$ error on the data qubits and to correct any single-qubit error on the auxiliary qubits.  The original auxiliary fermion approach is not especially efficient in correcting errors which occur on the data qubits. 

%We construct such mappings where the stabilizer operators can also be used for correcting/detecting single-qubit errors. Each site is labeled by the index $j_{x, y} = x + \length y$, where $x,\, y = 0,\,1,\ldots, \length-1$ are the horizontal and vertical coordinates of the site. We associate two data Majorana operators $\ff_{2j}$ and $\ff_{2j+1}$ and two auxiliary Majorana operators $\gamma_{2j}$ and $\gamma_{2j+1}$ to each lattice site. 

%Instead of using the JW transformation therein, we map 4 data Majorana operators and 4 auxiliary Majorana operators to Pauli operators of 4 qubits. This allows us to construct a code that corrects more than 90\% of single-qubit errors and detects the remaining single-qubit errors.

\section{Combining two unit cells in BVC}
\label{sec:combine_unit_cells}

In the last example, the Pauli operators of the physical qubits are expressed in terms of the Majorana fermion operators of the data modes and the auxiliary modes.  Here, we use the same gauge operators defined in Eq.~\eqref{eq:gauge_ops}, but we encode four data Majorana modes and four auxiliary Majorana modes into a block of 4 physical qubits.  This allows for correcting 10 out of 12 single-qubit errors in the block and detecting the remaining two. 

For simplicity of notation, we omit the block indices in the operators, e.g., $\ff_{2k+2}$ is denoted as $\ff_2$.  The Pauli operators on the $k$th block are constructed from the Majorana operators,
\begin{align}
 &\XX_0 = i\parity\ff_0 \ff_1\gamma_1 \,,\quad \YY_0 = i\parity\ff_2\ff_3\gamma_0\,, \label{eq:qubit0}\\[3pt]
 &\XX_1 = i\parity\ff_1\ff_2\gamma_{3}\,,\quad \YY_1 = i\parity\ff_0\ff_3\gamma_{2} \label{eq:qubit1}\,,\\[3pt]
  &\XX_2 = i\parity\ff_2\gamma_1\gamma_2\,,\quad \YY_2 = i\parity\ff_0\gamma_0\gamma_3\label{eq:qubit2}\,,\\[3pt]
 &\XX_3 = i\parity\ff_3\gamma_1\gamma_3\,,\quad \YY_3 = i\parity\ff_1\gamma_0\gamma_2 \label{eq:qubit3}\,,
\end{align}
where $P$ is a shorthand of the parity operator $P_0^{k-1}$, i.e., the product of all the preceding Majorana operators.  These operators commute with any single-mode Majorana operator on the first $k-1$ blocks and the corresponding Pauli operators.  These operators satisfy the commutation relations of the Pauli matrices. 
%The Pauli-$Z$ operators corresponding to Eq.~(\ref{eq:qubit0}-\ref{eq:qubit3}) are
%\begin{align}
% &\ZZ_0 = -iX_0Y_0 = -i\ff_0\ff_1\ff_2\ff_3\gamma_0\gamma_1 \,,\\[3pt]
% &\ZZ_1 = -iX_1Y_1 =- i\ff_0\ff_1\ff_2\ff_3\gamma_2\gamma_3 \,,\\[3pt]
% &\ZZ_2 = -iX_2Y_2 = -i\ff_0\ff_2\gamma_0\gamma_1\gamma_2\gamma_3\,,\\[3pt]
% &\ZZ_3 = -iX_3Y_3 = i\ff_1 \ff_3\gamma_0\gamma_1\gamma_2\gamma_3\,.
%\end{align}
Inversely, the Majorana operators take the form
\begin{align}
& \ff_0 = \parity\XX_0\YY_1\YY_2\ZZ_3\,, &&\gamma_0 = \parity\YY_0\ZZ_1\YY_2\YY_3\,, \label{eq:inverse_relation_a} \\
& \ff_1 = -\parity\XX_0\XX_1\ZZ_2\YY_3\,, &&\gamma_1 = -\parity\XX_0\ZZ_1\XX_2\XX_3\,,\\
& \ff_2 = \parity\YY_0\XX_1\XX_2\ZZ_3\,, &&\gamma_2 = -\parity\ZZ_0\YY_1\XX_2\YY_3\,,\\
& \ff_3 = -\parity\YY_0\YY_1\ZZ_2\XX_3\,, &&\gamma_3 = -\parity\ZZ_0\XX_1\YY_2\XX_3\,.\label{eq:inverse_relation_d}
\end{align}
The logical fermionic operators within the cell are
\begin{align}
 &i\ff_0\ff_3 = -\ZZ_0\XX_2 \YY_3 \,, && i\ff_1\ff_2 = -\ZZ_0\YY_2 \XX_3 \,,\\
 &i \ff_0 \ff_1 =  -\ZZ_1\XX_2  \XX_3\,, &&i \ff_2 \ff_3 =  \ZZ_1\YY_2  \YY_3
\end{align}
which correspond to the horizontal hopping terms and the occupation terms.   The inter-block horizontal hopping terms can be constructed by using the inverse relations~(\ref{eq:inverse_relation_a}-\ref{eq:inverse_relation_d}); the weight of the corresponding logical operators is seven.   All the terms containing one data Majorana operator and one auxiliary Majorana operator from the same block are three local, e.g.,
\begin{align}
 i\ff_0\gamma_1 =\XX_1 \ZZ_2 \YY_3 \,, \quad i\ff_1\gamma_1 = \YY_1\YY_2\ZZ_3\,.
\end{align}
Vertical hopping terms can be implemented by a 6-local Pauli operator corresponding to the product of two pairs of such operators.  The stabilizer can be either 6 local or 14 local depending whether they are intra-block or inter-blocks. 

There are 4 auxiliary Majorana operators $\gamma_{0,1,2,3}$ associated with a block, each of which form a gauge operator with another auxiliary Majorana operator from the block above or below.  To analyze the syndromes of the single-qubit errors, we first look at the parity term $\parity_0^{k-1}$.  The parity term anticommutes with gauge operators that share one common single-mode Majorana operator with it.  This set includes two kinds of gauge operators:
\begin{inparaenum}
\item those consist of one auxiliary Majorana operator to the left (right) of the $k$th block and an auxiliary Majorana operator below it,
\item those consist of one auxiliary Majorana operator to the right (left) of the $k$th block, including the $k$th block, and an auxiliary Majorana operator above it.
\end{inparaenum}
One can determine the block $k$, based on the syndromes of $P$, if a Pauli-$X$ or $Y$ error has occurred.  Any single-qubit Pauli-$X$ or $Y$ error contains a unique combination of the auxiliary Majorana operators in the block $k$, see Eqs.~\eqref{eq:qubit0}-\eqref{eq:qubit3}.  Therefore, their syndromes are different.

The Pauli-$Z$ errors do not involve the parity term.  For example, the Pauli operator $Z_0$ and $Z_1$ contain the auxiliary Majorana operators $\gamma_0\gamma_1$ and $\gamma_2\gamma_3$ in the block. These syndromes allows one to identify the block where the error has happened.  Both of the Pauli errors $Z_2$ and $Z_3$ contain the same product $\gamma_0 \gamma_1\gamma_2\gamma_3$ and have the same syndrome.   Therefore, the code can correct 10 out of the 12 single-qubit errors in the block and detect the remaining two single-qubit errors $Z_2$ and $Z_3$. 
% We call $\gamma_0$ and $\gamma_2$ upward auxiliary Majorana operators since they form gauge operators with the auxiliary operators from above.  Similarly, we call $\gamma_1$ and $\gamma_3$ downward auxiliary Majorana operators.  

\section{Constructing general MLSC codes}
\label{sec:general_mlsc}

In this appendix, we introduce a systematic procedure to construct general MLSC codes on a graph $G$.  Geometric locality is inevitably lost in the most general settings; we comment on the conditions when geometric locality can be retained.  Starting from a particular MLSC code, e.g., the BKSF, we modify the logical operators without changing the commutation relations~\eqref{eq:commu_quadratic_majo}.  Depending on whether the stabilizer group is changed, we discuss two different methods to generalize MLSCs.  Combining these two methods, one can construct arbitrary MLSCs.

In the first method, we introduce a new set of logical operators that also belong to the centralizer of the stabilizer group $\mathcal C(\mathcal S)$, i.e., the stabilizer group remains the same.  This can be achieved by a transformation that retains the standard anticommutation relations of the Majorana operators, similar to applying unitaries in the Clifford group to Pauli operators.  Any multi-mode Majorana operator (up to a sign) acting on $n$ fermionic modes can be represented by a $2n$-dimensional vector over \GFT.  For example, the Majorana operator $i\ff_0\ff_1\ff_2$ acting on two fermonic modes can be represented (up to the overall phase factor $i$) by the vector $(1 \; 1 \; 1 \; 0)$, where $1$ ($0$) indicates the presence (absence) of the corresponding single-mode Majorana operator in the product.  We introduce a matrix $M$, whose rows correspond to the new Majorana operators in terms of the old ones.  A specific example of $M$ for $n=2$ takes the form
\begin{align}\label{eq:4majo_trans}
    M_4 = \begin{pmatrix}
    1 & 1 & 1 & 0\\
    0 & 1 & 1 & 1\\
    1 & 0 & 1 & 1\\
    1 & 1 & 0 & 1
    \end{pmatrix}\,,
\end{align}
where the new Majorana operators satisfy the standard anticommutation relations~\eqref{eq:anti_commut}.  Since the overall parity is conserved by any quadratic Majorana operator, the Hamming weights of all the row vectors in $M$ have to be all odd or all even.  The four-Majorana transformation~\eqref{eq:4majo_trans} plays a similar role as the two-qubit gate in the Clifford group, whereas permutations of the Majorana modes are similar to single-qubit Clifford operations.  A general transformation matrix $M$ on $2n$ Majorana modes can be decomposed into a sequence of the four-Majorana transformation $M_4$ and row permutation matrices.  Any two rows of $M$ are orthogonal to each other due to the anticommutation relations of Majorana operators~\eqref{eq:anti_commut}.  To retain geometric locality, one can restrict themself to transformations between neighboring Majorana modes. 

In the second method, we modify the logical operators and then update the stabilizer group based on the modified logical operators.  A general procedure is:
\begin{enumerate}
    \item Find a Pauli operator $\Delta$ that anticommutes with at least one of the stabilizer operators, say $S$.
    \item Chose a logical operator $L$ and update it $L\mapsto L\Delta$,
    \item Find the set of logical operators $\{L'\}$, $L$ excluded, that anticommute with $\Delta$.
    \item Update these logical operators by multiplying them with the stabilizer operator $L'\mapsto L' S$.
    \item Update the stabilizer operators using the new logical operators based on Eq.~\eqref{eq:closed_path_relation}.
    \item Repeat the procedure for the remaining logical operators.
\end{enumerate}
This procedure produces a new MLSC with modified logical operators and stabilizers.  In order for the resulting MLSC code to be geometrically local, the choice of $\Delta$ should also be geometrically local.

\section{Algorithms to set initial qubit values}
\label{sec:initial_prep}

In this appendix, we give an algorithm to assign initial values to the qubits for preparation of a Slater determinant.  This algorithm requires neither a Hamiltonian path nor a spanning tree.
\begin{algorithm}[H]
\caption{Assign values to $z_{jk}$}\label{alg:initial_prod}
\begin{algorithmic}[1]
\Require
 \State $G$ \Comment a connected undirected graph
 \State $V$ \Comment the set of vertices of $G$
 \State $E$ \Comment the set of edges of $G$
\State $z_k\in\{\pm 1\}$ \Comment for each vertex $k\in V$
\begin{spacing}{0.5}
\Statex
\end{spacing}
\Procedure{RmLeg\ssp}{$k$}\;\;   \Comment remove the leg with the degree-one vertex $k\in V$, i.e., $\textproc{Deg}\ssp(k)=1$
\State find $j\in V$ such that $(j,k) \in E$ 
\State $z_{jk} \gets z_k$
\State $V \gets V \backslash \{k\}$
\State $E \gets E \backslash \{(j,k)\}$
\If {$\textproc{Deg}\ssp(j)=0$}
$V \gets V \backslash \{j\}$
\ElsIf {$\textproc{Deg}\ssp(j)=1$} 
$\textproc{RmLeg}\ssp(j)$
\EndIf
\EndProcedure
\While{there exist degree-one vertex $k\in V$}
\State $\textproc{RmLeg}\ssp (k)$
\EndWhile
\While{$E\neq \emptyset$}
\State randomly pick an edge $(j,k)\in E$
\State $z_{jk} \gets +1$
\State $E \gets E \backslash \{(j,k)\}$
\If {$\textproc{Deg}\ssp(k)=1$} $\textproc{RmLeg}\, (k)$
\EndIf
\If {$\textproc{Deg}\ssp(j)=1$}
 $\textproc{RmLeg}\ssp (j)$
\EndIf
\EndWhile
\end{algorithmic}
\end{algorithm}

\section{Code on single open plaquette}
\label{sec:single_plaquette}
\subsection{BKSF}
Here we work out the simple example of the BKSF stabilizer in a $2 \times 2$ graph of vertices, labeled $i$, $j$, $k$, $l$.  In this case, there are $4$ vertices, and $4$ edges representing the physical qubits.  There is one stabilizer following the general relation $N_E - N_V + 1$ for independent closed paths on the graph.  That reduces the total logical space to $3$ qubits, representing a parity conserving subspace.  A number of different stabilizers can result from different edge orderings, however the most convenient for our purposes follows a cycle such that the stabilizer is
\begin{align}
    S_{ijkl} &= \tilde\xi_{ij} \tilde\xi_{jk} \tilde\xi_{kl} \tilde\xi_{li} \notag \\
    &= (X_{ij} Z_{jk}) (X_{jk} Z_{kl})( X_{kl} Z_{li})( X_{li} Z_{ij}) \notag \\
    &= (X_{ij} Z_{ij}) (Z_{jk} X_{jk}) (Z_{kl} X_{kl}) (Z_{li} X_{li}) \notag \\
    &= -Y_{ij} Y_{jk} Y_{kl} Y_{li}
\end{align}
From this, we see that any individual $X$ or $Z$ error on a physical qubit of this code anti-commutes with this stabilizer.  As a result, it's possible to detect and post-select on any single qubit $X$ or $Z$ error, and more generally any odd number of $X$ or $Z$ errors. However it is easy to see that the code fails to detect a $Y$ error on any single qubit and by extension we cannot correct arbitrary single qubit errors with this encoding.

\section{Code on BKSF infinite lattice}
\label{sec:bksf_infinite_code}
Here we consider the properties of a BKSF encoding of a square regular lattice of fermions that extends in each direction.  For this code, the physical qubits on an edge always participate in two generating plaquettes, and each edge is treated identically to all others from symmetry.

The following $4$ stabilizers involve vertex $m$, and can be read directly from Fig \ref{fig:bk_plaquette} as
\begin{align}
S_{ijml} &= - X_{ij} Y_{jm} Y_{ml} X_{li} Z_{ii'} Z_{ii''} \notag \\
S_{jknm} &= - X_{jk} Y_{kn} Y_{nm} X_{jm} Z_{ij} Z_{jj''} \notag \\
S_{lmpo} &= - X_{lm} Y_{mp} Y_{op} X_{ol} Z_{ll'} Z_{li} \notag \\
S_{mnqp} &= - X_{mn} Y_{nq} Y_{qp} X_{pm} Z_{mj} Z_{ml}
\end{align}
where we have labeled the off figure indices by $i'$ for ones to the left of $i$ and $i''$ for ones above $i$.  These $4$ plaquettes constitute the complete set of stabilizers that act on the physical edge qubits $E_{mj}$ and $E_{ml}$.  We characterize the syndromes of each of these errors in Table \ref{fig:bksf_syndrome_inf}.  

To determine the possibility of error correction, one should verify that for any $\mathcal{E}_a^\dagger \mathcal{E}_b$ where $\mathcal{E}_p \in \{X_{mj}, Y_{mj}, Z_{mj}, X_{ml}, Y_{ml}, Z_{ml} \}$ there is a stabilizer operator that anti-commmutes with it.  This is sufficient for a recovery operation to exist for that set of errors.  From Table \ref{fig:bksf_syndrome_inf}, we see this is not the case for all errors.  Hence the BKSF transformation on an infinite square lattice with the given ordering is capable of detecting single qubit errors, but not correcting them in general.
\begin{figure}
\begin{center}
\begin{tabular}{c | c c c c}
   & $S_{ijml}$ & $S_{jknm}$ & $S_{lmpo}$ & $S_{mnqp}$ \\
  \hline			
  $X_{mj}$ & $-1$ & $1$ & $1$ & $-1$ \\
  $Y_{mj}$ & $1$ & $-1$ & $1$ & $-1$ \\
  $Z_{mj}$ & $-1$ & $-1$ & $1$ & $1$  \\
  $X_{ml}$ & $-1$ & $1$ & $1$ & $-1$ \\
  $Y_{ml}$ & $1$ & $1$ & $-1$ & $-1$ \\
  $Z_{ml}$ & $-1$ & $1$ & $-1$ & $1$ \\
  \hline
  $X_{mj} Y_{mj}$ & $-1$ & $-1$ & $1$ & $1$\\
  $X_{mj} Z_{mj}$ & $1$ & $-1$ & $1$ & $-1$ \\
  $X_{ml} Y_{ml}$& $-1$ & $1$ & $-1$& $1$\\
  $X_{ml} Z_{ml}$ & $1$ & $1$ & $-1$ & $-1$ \\
  $X_{mj} X_{ml}$ & $1$ & $1$ & $1$ & $1$ \\
  $X_{mj} Y_{ml}$ & $-1$ & $1$ & $-1$ & $1$ \\
  $X_{mj} Z_{ml}$ & $1$ & $1$ & $-1$ & $-1$\\
  $X_{ml} Y_{mj}$ & $-1$ & $-1$ & $1$ & $1$\\
  $Z_{ml} Y_{mj}$ & $-1$ & $-1$ & $-1$ & $-1$ \\
  $X_{ml} Z_{mj}$ & $1$ & $-1$ & $1$ & $-1$ \\
  $Z_{mj} Y_{ml}$ & $-1$ & $-1$ & $-1$ & $-1$ \\
  $Z_{mj} Z_{ml}$ & $1$ & $-1$ & $-1$ & $1$
\end{tabular}
\end{center}
\caption{Syndromes of different errors on the infinite square BKSF lattice.  The convention here currently uses $-1$ to imply anti-commutation with the error and $1$ to imply commutation. The first block looks at syndromes of single errors on the edges for the purpose of detection and the second block examines the related quantity $\mathcal{E}_a^\dagger \mathcal{E}_b$ related to the possibility of correcting those errors.  \label{fig:bksf_syndrome_inf}}
\end{figure}

\section{Code on BKSF torus}
\label{sec:bksf_torus_code}
Here we look at a patch of the BKSF code under periodic boundary conditions so that it forms a torus.  For a lattices of size $3 \times 3$ or larger, symmetries suggest it is again acceptable to look at the representative stabilizers that involve the vertex $m$ in the figure pictured in the main text.  From the infinite lattice, we can get the corresponding stabilizers from appropriate assignment of the indices with periodic boundary conditions corresponding to a 2D-torus.  This yields the stabilizers involving vertex $m$ as
\begin{align}
S_{ijml} &= - X_{ij} Y_{jm} Y_{ml} X_{li} Z_{ik} Z_{io} \notag \\
S_{jknm} &= - X_{jk} Y_{kn} Y_{nm} X_{jm} Z_{ij} Z_{jp} \notag \\
S_{lmpo} &= - X_{lm} Y_{mp} Y_{op} X_{ol} Z_{ln} Z_{li} \notag \\
S_{mnqp} &= - X_{mn} Y_{nq} Y_{qp} X_{pm} Z_{mj} Z_{ml}
\end{align}
which from inspection we see are qualitatively the same as the case for the infinite lattice.  That is, we expect the same commutation relations for the stabilizers and the representative edges $E_{ml}$ and $E_{mj}$.  

\section{Code on BKSF open section}
\label{sec:open_section}
Considering now a finite square lattice with open boundary conditions with the standard ordering in this work, we may again work from Fig \ref{fig:bk_plaquette}.  In this case, all the plaquettes except those on the left or upper boundaries of the graph are the same, where the dangling $Z$ operators are truncated.

For interior edges more than 1 plaquette from a boundary, the same detection and correction results from the infinite case apply.  However, at the boundary, there are generally edges that are covered by only 1 or 2 plaquettes instead of 3, as in the interior cases.  As a result, even error detection at the edges can become problematic.  For example, in the bottom-left plaquette of a square lattice with open boundary conditions, one is incapable of detecting a single $X$ error on that edge.  This means that we expect there are a number of undetectable errors in this case as well, especially at boundaries.  This may still allow for the mitigation of some errors, but is less powerful than the periodic case for this choice of ordering.  For a finite graph, it may be possible to devise a custom or non-uniform ordering that restores the power to detect errors, however we do not investigate that here.

\end{document}